\documentclass[prd,showpacs,showkeys,nofootinbib,twocolumn]{revtex4-1}

\usepackage{graphicx}
\usepackage{amsmath,amsfonts,amssymb,amsxtra,latexsym}

\usepackage{hyperref}
\hypersetup{
    colorlinks=true,
    linkcolor=red,
    citecolor=blue,      
    urlcolor=blue,
}

\renewcommand{\pre}[2]{{}^{(#2)}#1}
\newcommand{\C}{\bar{C}}
\allowdisplaybreaks

\begin{document}

\title{Extended gravitational clock compass: new exact solutions and simulations}

\author{Gerald Neumann}
\email{geneumann@udec.cl}
\affiliation{Departamento de F\'isica, Universidad de Concepci\'on, Casilla 160-C, Concepci\'on, Chile} 

\author{Dirk Puetzfeld}
\email{dirk.puetzfeld@zarm.uni-bremen.de}
\homepage{http://puetzfeld.org}
\affiliation{University of Bremen, Center of Applied Space Technology and Microgravity (ZARM), 28359 Bremen, Germany} 

\author{Guillermo F. Rubilar}
\email{grubilar@udec.cl}
\affiliation{Departamento de F\'isica, Universidad de Concepci\'on, Casilla 160-C, Concepci\'on, Chile{}} 

\date{ \today}

\begin{abstract}
By extending the framework of the gravitational clock compass we show how a suitably prepared set of clocks can be used to extract information about the gravitational field in the context of General Relativity. Conceptual differences between the extended and the standard clock compass are highlighted. Particular attention is paid to the influence of kinematic quantities on the measurement process and the setup of the compass. Additionally, we present results of simulations of the inference process for the acceleration and the curvature components. Several examples of different strategies for the computation of the posterior probability distributions of the curvature components are discussed. This allows us to anticipate the precision with which physical quantities could be determined in a realistic measurement.
\end{abstract}


\maketitle


\section{Introduction}\label{introduction_sec}

The question of how the gravitational field can be determined in an operational way is of fundamental importance in gravitational physics. One particular approach which has received attention in recent years is the so-called \textit{gravitational clock compass} \cite{Puetzfeld:etal:2018:2}, which adapts the original idea of Szekeres \cite{Szekeres:1965}, which in turn may be viewed as an implementation of the concepts introduced by Pirani \cite{Pirani:1956} and Synge \cite{Synge:1960} in the context of the geodesic deviation equation. Here we make use of clock measurements in order to determine the curvature of spacetime. This method has been developed in \cite{Puetzfeld:etal:2018:2,Obukhov:Puetzfeld:2019} and can be viewed as complementary to the use of deviation equations \cite{Puetzfeld:Obukhov:2016:1} and swarms of test bodies. An alternative derivation of the clock compass and its use in the context of exact gravitational wave spacetimes can be found in \cite{Hogan:Puetzfeld:2020:1}. 

While the aforementioned works established the foundations of the clock compass and demonstrated its ability to measure all components of the gravitational field in special, as well as in general spacetimes, the present work is focused on methods which will help to manifest its practical use in future experiments. To achieve this goal, we extend the framework of the gravitational clock compass in two ways. 

First, we derive new analytical solutions which determine the acceleration and angular velocity of the reference frame in which the measurements are performed. The same is also carried out for the 20 independent components of the curvature tensor of General Relativity. Conceptual differences between the extended and the standard clock compass are highlighted. Particular attention is paid to the influence of kinematic quantities on the measurement process and the optimal setup of the compass.

Second, we explore the way in which the determination of the physical quantities (acceleration, angular velocity, curvature) could actually be performed from the data collected by a given configuration of clocks. We do this by generating mock data sets which are in turn used in the simulation of the parameter determination process. Special focus is put on different admissible strategies to compute each curvature component. 

The structure of the paper is as follows: In section \ref{sec_general_curved_case} we review the frequency ratio of moving clocks in a general spacetime background. This is followed by an introduction to the gravitational clock compass in section \ref{sec_clock_compass}. Subsequently we show in section \ref{sec_determination_state_of_motion} how a clock compass can be used to determine the state of motion (acceleration and angular velocity) of a reference frame with respect to a free-falling frame. In section \ref{sec_curvature_determination} the derivation of a general exact solution for the curvature components in terms of the measurable frequency ratios, as well as the position and velocities of the clocks is presented. Subsequently we discuss in \ref{types_of_measurements_sec} how, and with which precision, the different parameters could be determined, by simulating the parameter estimation using mock data. In section \ref{sec_data_gen} we explain how we generate our mock data, and the incorporation of measurement errors into the data. The mock data is then used to determine probability distributions for the kinematic quantities, as well as for the gravitational field components, by means of the implementation of a MCMC method. Particular examples for the determination of the acceleration are presented in section \ref{sec_a}. We then proceed to discuss in detail the different strategies for the determination of curvature components in section \ref{error-curv}. Our conclusions are discussed in section \ref{sec_conclusions}. Finally, we collect some useful complementary material in the appendices \ref{sec_notation}--\ref{app_RB}.

\section{Frequency ratio of clocks}\label{sec_general_curved_case}

We start with the general result of \cite{Puetzfeld:etal:2018:2} for the ratio of the proper times $ds|_X$ and $ds|_Y$ of two clocks $X$ and $Y$, respectively; the former at a position $y^\alpha$ and the latter at the origin of a Fermi coordinate system. The frame associated to this coordinate system is characterized by the acceleration $a_\alpha$ and the angular velocity $\omega_\alpha$, which in this work we assume to be time-independent. Then the frequency ratio between two clocks, see \cite{Puetzfeld:etal:2018:2} for more details, takes the following form:
\begin{align}
\left(\frac{ds|_X}{ds|_Y}\right)^2 =&\left(\frac{dy^{0}}{ds|_Y}\right)^2 \Bigr[ 1 - \delta_{\alpha \beta} v^\alpha v^\beta + 2 a_\alpha y^\alpha \nonumber \\
& + y^\alpha y^\beta \left( a_\alpha a_\beta - \delta_{\alpha \beta} \omega_\gamma \omega^\gamma + \omega_\alpha \omega_\beta - R_{0 \alpha \beta 0} \right) \nonumber \\
&+ 2 v^\alpha \varepsilon_{\alpha \beta \gamma} y^\beta \omega^\gamma - \frac{4}{3} v^\alpha y^\beta y^\gamma R_{ \alpha \beta \gamma 0} \nonumber \\
& - \frac{1}{3} v^\alpha v^\beta y^\gamma y^\delta R_{\gamma \alpha \beta \delta} \Bigr] + {\mathcal O}(3) , \label{general_freq_ratio_definition}
\end{align}
were we define the auxiliary function $\C$ so that
\begin{align}
\C \left( y^\alpha, v^\alpha , a^\alpha, \omega^\alpha, R_{abcd} \right) + 1 := \left(\frac{ds|_X}{ds|_Y}\right)^2.  \label{shortcut}
\end{align}
Our conventions and notation are summarized in appendix \ref{sec_notation}.

In contrast to \cite{Puetzfeld:etal:2018:2}, we will assume that the reference clock is always at rest w.r.t.\ the reference frame. Then, $(dy^{0})/(ds|_Y) = 1$ and, together with \eqref{general_freq_ratio_definition}, they imply that
\begin{align}
\C &=\Bigr[ - \delta_{\alpha \beta} v^\alpha v^\beta + 2 a_\alpha y^\alpha  \nonumber \\
&\quad + y^\alpha y^\beta \left( a_\alpha a_\beta - \delta_{\alpha \beta} \omega_\gamma \omega^\gamma + \omega_\alpha \omega_\beta - R_{0 \alpha \beta 0} \right) \nonumber \\
&\quad + 2 v^\alpha \varepsilon_{\alpha \beta \gamma} y^\beta \omega^\gamma  - \frac{4}{3} v^\alpha y^\beta y^\gamma R_{ \alpha \beta \gamma 0} \nonumber \\
&\quad - \frac{1}{3} v^\alpha v^\beta y^\gamma y^\delta R_{\gamma \alpha \beta \delta} \Bigr] + {\mathcal O}(3) . \label{general_freq_ratio_definition_our}
\end{align}

The ratio $\C$ is related to the redshift $z$ of $X$ w.r.t.\ $Y$, since by definition
\begin{align}
 1 + z := \left(\frac{ds|_Y}{ds|_X}\right), 
\end{align}
thus
\begin{align}
 \C + 1 = \left(\frac{ds|_X}{ds|_Y}\right)^2 = (1 + z)^{-2}. \label{meaning_of_c}
\end{align}
If $z \ll 1$, then 
\begin{align}
 \C \approx - 2z. \label{meaning_of_c_approx}
\end{align}

Fermi normal coordinates can be thought of as the natural extension of inertial Cartesian coordinates \cite{Synge:1960}. One should note, that the validity of the coordinate system in the vicinity of the central observer, is one of the limiting factors of the whole framework. By construction, the coordinate system used here is valid to describe physical phenomena in a small region around the world line of the central observer. The corresponding domain depends on the actual state of motion, in particular on the magnitudes of the acceleration, angular velocity and the curvature of spacetime, since they define corresponding distance scales: $\ell_{\rm accel} = c^2/|a^\alpha|$, $\ell_{\rm rot} = c/|\omega^\alpha|$ and $\ell_{\rm curv} = \min \{ |R_{abcd}|^{-1/2}, |R_{abcd}| / |\partial_e R_{abcd}| \}$ \cite{Ni:Zimmermann:1978}. As in the previous works on the gravitational compass \cite{Puetzfeld:etal:2018:2,Szekeres:1965,Obukhov:Puetzfeld:2019,Hogan:Puetzfeld:2020:1}, we make the implicit assumption that the coordinate system is of a sufficient accuracy w.r.t.\ the effects to be measured, without actually specifying the details of the field we are interested in. For a region near the surface of the Earth the most severe restriction is given by the curvature distance scale, since $\ell_{\rm grav} = \min \{ |r_{\rm s}/r^3|^{-1/2}, \  r/3 \}\approx r_{\rm Earth} / 3 \approx 10^6 {\rm m} $. We take this restriction into account in the choice of the distances considered in the examples shown in Sect. \ref{types_of_measurements_sec}. Depending on the desired level of accuracy, and spacing of the clocks, a very accurate modeling of the gravitational field between the clocks would be required. It is also clear, that a modeling of the time transfer up to any required order can be performed in an iterative fashion, see for example \cite{Poncin-Lafitte:etal:2004,Teyssandier:etal:2008:1,Teyssandier:etal:2008:2} for a covariant framework in terms of the world function.

\section{Clock compass setup}\label{sec_clock_compass}

\begin{figure}
\begin{center}
\includegraphics[scale = 0.4]{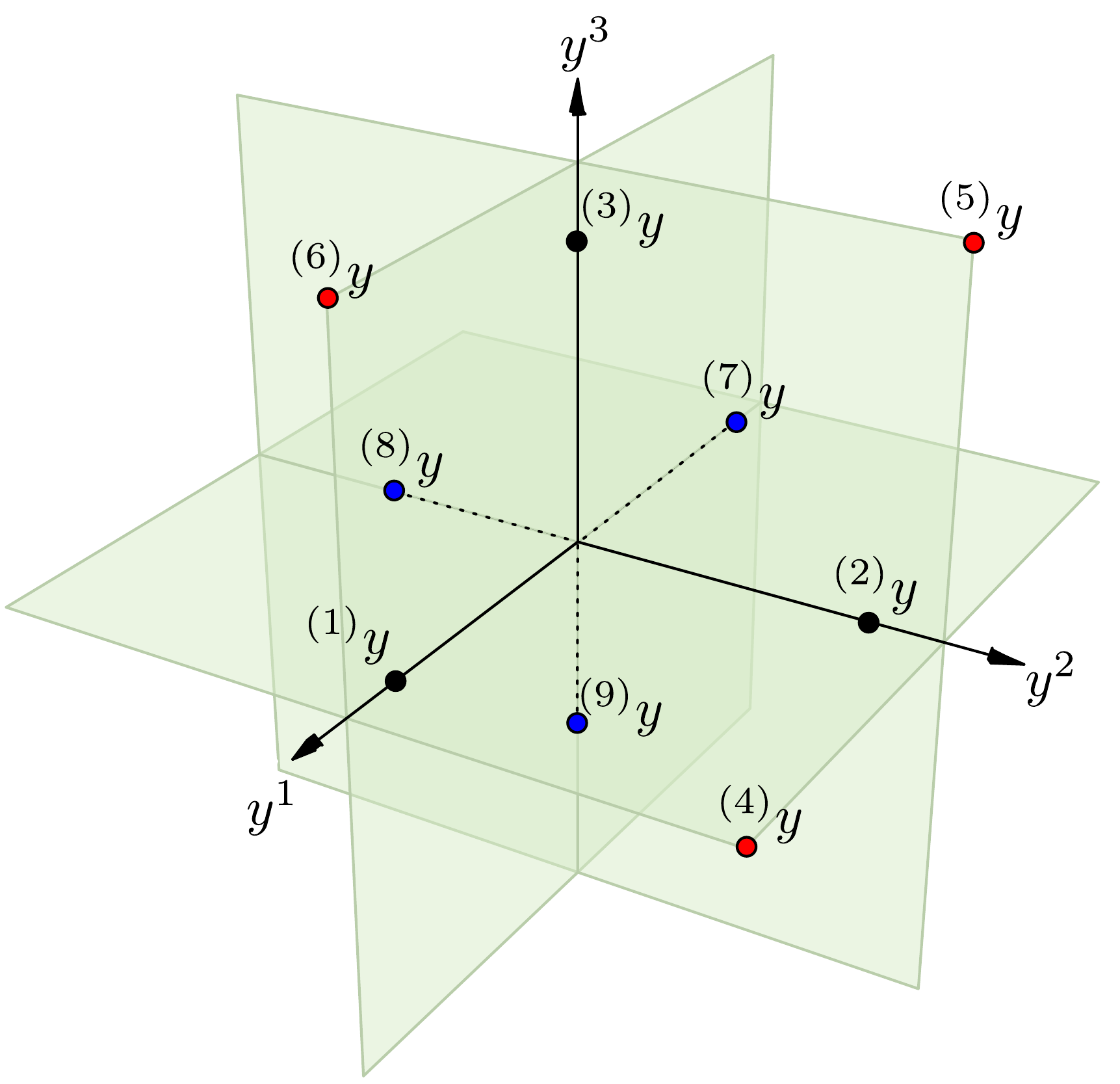}
\end{center}
\caption{\label{pos_1} Positions of the 9 clocks chosen for the first clock array, see Eq.\ \eqref{position_setup}.}
\end{figure}

For the location of the clocks considered in the present work, we use the same type of arrangement as in \cite{Puetzfeld:etal:2018:2}, but here we allow for numerical values different from $1$ for the positions of the clocks. Thereby their distance w.r.t.\ the central reference clock appears explicitly in the equations, which will turn out to be useful in the modeling of the measurement process. In particular, we will study how the precision of the determination of the physical parameters depends on the distance of the clocks to the reference world line $Y$. We start by labeling 9 different initial values for the positions of the clocks, as follows:
\begin{align}
{}^{(1)}y^{\alpha}&=\left(\begin{array}{c} y_{11} \\ 0\\ 0\\ \end{array} \right), 
{}^{(2)}y^{\alpha}=\left(\begin{array}{c} 0 \\ y_{22}\\ 0\\ \end{array} \right), 
{}^{(3)}y^{\alpha}=\left(\begin{array}{c} 0 \\ 0\\ y_{33}\\ \end{array} \right), \nonumber \\
{}^{(4)}y^{\alpha}&=\left(\begin{array}{c} y_{41}\\ y_{42}\\ 0\\ \end{array} \right),
{}^{(5)}y^{\alpha}=\left(\begin{array}{c}  0\\ y_{52}\\ y_{53}\\ \end{array} \right),
{}^{(6)}y^{\alpha}=\left(\begin{array}{c}  y_{61}\\ 0\\ y_{63}\\ \end{array} \right), \label{position_setup} \\
{}^{(7)}y^{\alpha}&=-{}^{(1)}y^{\alpha}, \quad
{}^{(8)}y^{\alpha}=-{}^{(2)}y^{\alpha}, \quad 
{}^{(9)}y^{\alpha}=-{}^{(3)}y^{\alpha}. \nonumber 
\end{align}
These positions are sketched in Fig.\ \ref{pos_1}.

For the velocities, we consider the most general case in which each clock has arbitrary direction and speed, and write 
\begin{eqnarray}
&&{}^{(n)}v^{\alpha}=\left(\begin{array}{c} v_{n1} \\ v_{n2}\\ v_{n3}\\ \end{array} \right), \label{velocity_setup}
\end{eqnarray}
where $n$ denotes the $n$-th clock, $n=1,\dots,9$.  \\

Acceleration and angular velocity are properties of the reference system. In this sense, there is only one value for the vectors $a^\alpha$ and $\omega^\alpha$, given the choice of reference frame, and we denote them as
\begin{eqnarray}
a_{\alpha}=\left(\begin{array}{c} a_{1} \\ a_{2}\\ a_{3}\\ \end{array} \right), \ \text{ and } \ 
\omega_{\alpha}=\left(\begin{array}{c} \omega_{1} \\ \omega_{2}\\ \omega_{3}\\ \end{array} \right).  \label{rotaccel_setup}
\end{eqnarray}

\section{Determination of linear acceleration and angular velocity}\label{sec_determination_state_of_motion}

\subsection{Determination of linear acceleration}\label{paragraph_fermi_normal_flat_linear_accel_determination}

For the determination of the linear acceleration, we follow a very similar procedure as the one outlined in \cite{Puetzfeld:etal:2018:2}, and consider the simplest case in which the contribution of the curvature is neglected. We start by rearranging \eqref{general_freq_ratio_definition_our} as follows 
\begin{eqnarray}\label{2ay2aayyB}
2a_\alpha y^\alpha + a_\alpha a_\beta y^\alpha y^\beta = B_1(\C, y^\alpha, v^\alpha,\omega^\alpha),
\end{eqnarray} 
here all measured proper time ratios, as well as all prescribed quantities, are collected in the quantity $B_1$ on the r.h.s.\ of Eq.\ \eqref{2ay2aayyB}, which we define as:
\begin{eqnarray}
B_1(\C, y^\alpha, v^\alpha,\omega^\alpha) &:=& \C + v^2  - y^\alpha y^\beta \left( \omega_\alpha \omega_\beta - \delta_{\alpha \beta} \omega^2 \right) \nonumber \\
&&- 2 v^\alpha \varepsilon_{\alpha \beta \gamma}  y^\beta \omega^\gamma, \label{rhs_accel_determination}
\end{eqnarray}
where $v^2 := \delta_{\alpha\beta}v^\alpha v^\beta$ and $\omega^2 := \delta_{\alpha\beta}\omega^\alpha \omega^\beta$.

Taking into account expressions (\ref{position_setup})-(\ref{rotaccel_setup}) for each clock, we end up with the system
\begin{align}
2a_\alpha {}^{(n)}y^\alpha + a_\alpha a_\beta {}^{(n)}y^\alpha {}^{(n)}y^\beta &= B_1(\pre{\C}{n}, {}^{(n)}y^\alpha, {}^{(n)}v^\alpha, \omega^\alpha) \nonumber \\
&=: {}^{(n)}B_1,
\end{align} 
where $\pre{\C}{n}$ is the value of the function given by \eqref{general_freq_ratio_definition_our} evaluated at the position $\pre{y}{n}^\alpha$ and the velocity $\pre{v}{n}^\alpha$.

In order to determine the linear acceleration we use 3 pairs of clocks at opposite positions, namely (${}^{(1)}y^{\alpha}, {}^{(7)}y^{\alpha}$), (${}^{(2)}y^{\alpha},{}^{(8)}y^{\alpha}$) and (${}^{(3)}y^{\alpha},{}^{(9)}y^{\alpha}$). This yields a set of equations which can be used to solve for $a_\alpha$ leading, in terms of the $\C$'s, to
\begin{align}
\nonumber a_{1} =& \frac{1}{4 y_{11}} \left({}^{(1)}\C - {}^{(7)}\C - 2 \omega_{2} v_{1 3} y_{11} - 2 \omega_{2} v_{7 3} y_{11} \right.  \\
& \phantom{\frac{1}{4 y_{11}} \left( \right.} \left. + 2 \omega_{3} v_{1 2} y_{11} + 2 \omega_{3} v_{7 2} y_{11} + {}^{(1)}v^2 - {}^{(7)}v^2 \right), \nonumber \\ 
\nonumber a_{2} =& \frac{1}{4 y_{22}} \left({}^{(2)}\C - {}^{(8)}\C + 2 \omega_{1} v_{2 3} y_{22} + 2 \omega_{1} v_{8 3} y_{22} \right. \\
\nonumber & \phantom{\frac{1}{4 y_{11}} \left( \right.} \left. - 2 \omega_{3} v_{2 1} y_{22} - 2 \omega_{3} v_{8 1} y_{22} + {}^{(2)}v^2 - {}^{(8)}v^2\right), \\
\nonumber a_{3} =& \frac{1}{4 y_{33}} \left({}^{(3)}\C - {}^{(9)}\C - 2 \omega_{1} v_{3 2} y_{33} - 2 \omega_{1} v_{9 2} y_{33}  \right. \\
& \phantom{\frac{1}{4 y_{11}} \left( \right.} \left. + 2 \omega_{2} v_{3 1} y_{33} + 2 \omega_{2} v_{9 1} y_{33} + {}^{(3)}v^2 - {}^{(9)}v^2\right). \label{linaccel_solution}
\end{align}
These expressions allows us to compute the acceleration from an arrangement of 6 clocks with arbitrary velocities as parameterized in \eqref{velocity_setup}. If we consider the particular case in which each clock has the same velocity, i.e.\ ${}^{(n)}v^\alpha = (v_1,v_2,v_3)$, we obtain 
\begin{align}
 \nonumber a_{1} &= \frac{1}{4 y_{11}} \left({}^{(1)}\C - {}^{(7)}\C - 4 \omega_{2} v_{3} y_{11} + 4 \omega_{3} v_{2} y_{11} \right), \\ 
\nonumber a_{2} &= \frac{1}{4 y_{22}} \left({}^{(2)}\C - {}^{(8)}\C + 4 \omega_{1} v_{3} y_{22}- 4 \omega_{3} v_{1} y_{22}\right), \\
a_{3} &= \frac{1}{4 y_{33}} \left({}^{(3)}\C - {}^{(9)}\C - 4 \omega_{1} v_{2} y_{33}  + 4 \omega_{2} v_{1} y_{33} \right). \label{linaccel_solution_same_v}
\end{align}
We notice that in these expressions the acceleration $a_\alpha$ depends on the angular velocity $\omega_\alpha$. In \cite{Puetzfeld:etal:2018:2} the velocities of all clocks were chosen to be parallel to $\omega_\alpha$, an assumption not made in \eqref{linaccel_solution_same_v}. If those vectors are parallel, then the terms proportional to $\omega_\alpha$ vanish.

If we set the velocities of the clocks to zero, we obtain
\begin{align}
 \nonumber a_{1} &= \frac{1}{4 y_{11}} \left({}^{(1)}\C - {}^{(7)}\C  \right), \\ 
\nonumber a_{2} &= \frac{1}{4 y_{22}} \left({}^{(2)}\C - {}^{(8)}\C  \right), \\
a_{3} &= \frac{1}{4 y_{33}} \left({}^{(3)}\C - {}^{(9)}\C  \right). \label{linaccel_solution_zero_v}
\end{align}
This of course agrees with the results in \cite{Puetzfeld:etal:2018:2} if $c^2_{11} = 0$. 

\subsection{Determination of angular velocity}\label{paragraph_fermi_normal_flat_rotational_velo_determination}

Analogously to the strategy in the preceding section, we rearrange the system \eqref{general_freq_ratio_definition_our} as follows:
\begin{eqnarray}
 2 v^\alpha \varepsilon_{\alpha \beta \gamma}  y^\beta \omega^\gamma - y^\alpha y^\beta \left(\delta_{\alpha \beta} \omega^2 - \omega_\alpha \omega_\beta \right) =: B_2(y^\alpha, v^\alpha,a^\alpha),\nonumber \\
\end{eqnarray} 
where
\begin{eqnarray}
B_2(y^\alpha, v^\alpha, a^\alpha) &:=& \C + v^2  - 2a_\alpha y^\alpha - a_\alpha a_\beta y^\alpha y^\beta. \nonumber \\ \label{rhs_accel_determination_2}
\end{eqnarray}
Taking into account (\ref{position_setup})-(\ref{rotaccel_setup}) we end up with
\begin{multline}
 2{}^{(n)}v^\alpha \varepsilon_{\alpha \beta \gamma}  {}^{(n)}y^\beta \omega^\gamma - {}^{(n)}y^\alpha {}^{(n)}y^\beta \left(\delta_{\alpha \beta} \omega^2 -\omega_\alpha \omega_\beta \right) \\
 =  B_2({}^{(n)}y^\alpha, {}^{(n)}v^\alpha,a^\alpha) =: {}^{(n)}B_2 .
\end{multline}

\subsubsection{Same initial conditions as in \cite{Puetzfeld:etal:2018:2}}\label{omega-2order-1}

For reference, we first consider a configuration of six clocks with the same initial conditions as in \cite{Puetzfeld:etal:2018:2}, i.e.\ using clocks at the positions ${}^{(1)}y^{\alpha}$, ${}^{(2)}y^{\alpha}$ and ${}^{(3)}y^{\alpha}$, with velocities given by
\begin{eqnarray}
\nonumber {}^{(1)}v^{\alpha} &=& \left(\begin{array}{c} v_{11} \\ 0 \\ 0 \\ \end{array} \right), \quad {}^{(2)}v^{\alpha}=\left(\begin{array}{c} 0 \\ v_{22} \\ 0 \\ \end{array} \right),\\
\quad {}^{(3)}v^{\alpha} &=& \left(\begin{array}{c} 0 \\ 0\\ v_{33}\\ \end{array} \right). \label{rot_init}
\end{eqnarray} 
We denote by ${}^{(1,2)}\C$ the value of the function $\C$ given by \eqref{general_freq_ratio_definition_our} evaluated for the position ${}^{(1)}y^{\alpha}$ and velocity ${}^{(2)}v^{\alpha}$, etc. Then the angular velocity can be determined in terms of the values of ${}^{(1,1)}\C$, ${}^{(1,2)}\C$, ${}^{(2,2)}\C$, ${}^{(2,3)}\C$, ${}^{(3,1)}\C$ and ${}^{(3,3)}\C$ of each clock with corresponding position and velocity. We obtain:
\begin{eqnarray}
\omega_1&=&\frac{1}{2v_{33}y_{22}} \left[{}^{(2,2)}\C - {}^{(2,3)}\C + v_{22}^2 - v_{33}^2 \right], \nonumber \\
\omega_2&=&\frac{1}{2v_{11}y_{33}} \left[{}^{(3,3)}\C - {}^{(3,1)}\C + v_{33}^2 - v_{11}^2 \right], \nonumber \\
\omega_3&=&\frac{1}{2v_{22}y_{11}} \left[{}^{(1,1)}\C - {}^{(1,2)}\C + v_{11}^2 - v_{22}^2 \right]. \label{rotvel_solution_C} 
\end{eqnarray}
Note that, unlike the result \eqref{linaccel_solution} for $a^\alpha$, this solution for the angular velocity does not depend on the value of the acceleration of the frame. This is due to our choice for the positions and velocities of the present clock configuration, which leads to a set of equations in which the contribution of $a^\alpha$ cancels out. This behavior was also present in the analogous result reported in \cite{Puetzfeld:etal:2018:2}, although in that case the expression is different due to the different choice of the velocity of the reference clock.

\subsubsection{Velocity perpendicular to the position}

Now, we consider a configuration of clocks slightly different from the one shown in the previous section. We use pairs of clocks at the same positions as the previous section. These are given by ${}^{(1)}y^{\alpha}$, ${}^{(2)}y^{\alpha}$ and ${}^{(3)}y^{\alpha}$, with velocities perpendicular to their respective position vector, given by
\begin{align}
\nonumber {}^{(1)}v^{\alpha}&=\left(\begin{array}{c} 0 \\ v_{12}\\ 0\\ \end{array} \right), & {}^{(2)}v^{\alpha}&=\left(\begin{array}{c} 0 \\ 0\\ v_{23}\\ \end{array} \right), \\
\nonumber {}^{(3)}v^{\alpha}&=\left(\begin{array}{c} v_{31} \\ 0\\ 0\\ \end{array} \right), & {}^{(4)}v^{\alpha}&=\left(\begin{array}{c} 0\\ -v_{42}\\ 0\\ \end{array} \right), \\
{}^{(5)}v^{\alpha}&=\left(\begin{array}{c}  0\\ 0\\- v_{53}\\ \end{array} \right), & {}^{(6)}v^{\alpha}&=\left(\begin{array}{c}  -v_{61}\\ 0\\ 0\\ \end{array} \right). \label{velocity_setup_omega}
\end{align}

Considering this configuration, the solution for the angular velocity components can be written in terms of the $\C$'s as
\begin{align}
\nonumber \omega^1 &= \frac{-\pre{\C}{2,2} + \pre{\C}{2,5} - v^2_{23} + v^2_{53}}{2 y_{22} (v_{23} + v_{53} )}, \\
\nonumber \omega^2 &= \frac{-\pre{\C}{3,3} + \pre{\C}{3,6} - v^2_{31} + v^2_{61}}{2 y_{33} (v_{31} + v_{61} )}, \\
 \omega^3 &= \frac{-\pre{\C}{1,1} + \pre{\C}{1,4} - v^2_{12} + v^2_{42}}{2 y_{11} (v_{12} + v_{42} )}, \label{omega_uno}
\end{align}
where ${}^{(1,1)}\C$ is the value of the function $\C$ given by \eqref{general_freq_ratio_definition_our}, evaluated for the position ${}^{(1)}y^{\alpha}$ and velocity ${}^{(1)}v^{\alpha}$, etc. This solution can be simplified considering that we set the velocity of one of the clocks of the pair to zero, obtaining 
\begin{align}
\nonumber \omega^1 &= \frac{-\pre{\C}{2,2} + \pre{\C}{2,0} - v^2_{23} }{2 y_{22} v_{23} }, \\
\nonumber \omega^2 &= \frac{-\pre{\C}{3,3} + \pre{\C}{3,0} - v^2_{31} }{2 y_{33} v_{31} }, \\
 \omega^3 &= \frac{-\pre{\C}{1,1} + \pre{\C}{1,0} - v^2_{12} }{2 y_{11} v_{12} }, \label{omega_dos}
\end{align}
where ${}^{(1,0)}\C$ is the value of the function $\C$ given by \eqref{general_freq_ratio_definition_our} evaluated for the position ${}^{(1)}y^{\alpha}$ and zero velocity, etc. 

Alternatively, a simpler solution for this case can be found if we use 3 pairs of clocks at the same positions explained above, but choosing the velocities of each pair with the same position to be equal in magnitude but with opposite directions. Under these conditions, we obtain
\begin{align}
\nonumber \omega^1 &= \frac{-\pre{\C}{2,2} + \pre{\C}{2,-2} }{4 y_{22} v_{23} }, \\
\nonumber \omega^2 &= \frac{-\pre{\C}{3,3} + \pre{\C}{3,-3} }{4 y_{33} v_{31} }, \\
 \omega^3 &= \frac{-\pre{\C}{1,1} + \pre{\C}{1,-1} }{4 y_{11} v_{12} }, \label{omega_tres}
\end{align}
where ${}^{(1,-1)}\C$ is the value of the function $\C$ given by \eqref{general_freq_ratio_definition_our}, evaluated for the position ${}^{(1)}y^{\alpha}$ and velocity $-{}^{(1)}v^{\alpha}$, etc. 

\subsubsection{Clocks at rest}

For clocks at rest, it is interesting to notice that the absolute values of the components $\omega_\alpha$ can be determined by using only three clocks. Indeed, with the three clocks at the positions ${}^{(1)}y^{\alpha}$, ${}^{(2)}y^{\alpha}$ and ${}^{(3)}y^{\alpha}$ we can obtain, using Eq.\ \eqref{general_freq_ratio_definition_our}, expressions for ${}^{(1)}\C$, ${}^{(2)}\C$ and ${}^{(3)}\C$, which form a system of three equations for the three unknowns $\omega_{1}^{2}$, $\omega_{2}^{2}$ and $\omega_{3}^{2}$. The solution then is found to be:
\begin{align}
\omega_{1}^{2} &= \frac{{}^{(1)}\C}{2 y_{11}^{2}} - \frac{{}^{(2)}\C }{2 y_{22}^{2}} - \frac{{}^{(3)}\C}{2 y_{33}^{2}} \nonumber \\ 
&\quad - \frac{a_{1}^{2}}{2} + \frac{a_{2}^{2}}{2} + \frac{a_{3}^{2}}{2} - \frac{a_{1}}{y_{11}}  + \frac{a_{2}}{y_{22}}  + \frac{a_{3}}{y_{33}}, \\
\nonumber \omega_{2}^{2} &= - \frac{{}^{(1)}\C}{2 y_{11}^{2}} + \frac{{}^{(2)}\C}{2 y_{22}^{2}} - \frac{{}^{(3)}\C }{2 y_{33}^{2}} \\
&\quad + \frac{a_{1}^{2}}{2} - \frac{a_{2}^{2}}{2} + \frac{a_{3}^{2}}{2}   + \frac{a_{1}}{y_{11}} - \frac{a_{2}}{y_{22}} + \frac{a_{3}}{y_{33}}, \\
\nonumber \omega_{3}^{2} &= - \frac{{}^{(1)}\C}{2 y_{11}^{2}} - \frac{{}^{(2)}\C}{2 y_{22}^{2}} + \frac{{}^{(3)}\C}{2 y_{33}^{2}} \\
&\quad + \frac{a_{1}^{2}}{2}  + \frac{a_{2}^{2}}{2} - \frac{a_{3}^{2}}{2} + \frac{a_{1}}{y_{11}}  + \frac{a_{2}}{y_{22}}  - \frac{a_{3}}{y_{33}}.
\end{align}
If we position the three clocks at the same distance to the central reference clock, i.e.\ $y_{11} = y_{22} = y_{33} = y$, we obtain
\begin{align}
 \omega_{1}^{2} =& \frac{1}{2 y^{2}} \left({}^{(1)}\C - {}^{(2)}\C - {}^{(3)}\C  - a_{1}^{2} y^{2}  + a_{2}^{2} y^{2} + a_{3}^{2} y^{2} \right. \nonumber \\
& \phantom{\frac{1}{2 y^{2}}} - 2 a_{1} y + 2 a_{2} y  + 2 a_{3} y \Big), \\
 \omega_{2}^{2} =& \frac{1}{2 y^{2}} \left(- {}^{(1)}\C + {}^{(2)}\C - {}^{(3)}\C + a_{1}^{2} y^{2} - a_{2}^{2} y^{2} + a_{3}^{2} y^{2} \right. \nonumber \\
 & \phantom{\frac{1}{2 y^{2}}} + 2 a_{1} y - 2 a_{2} y  + 2 a_{3} y \Big), \\
 \omega_{3}^{2} =& \frac{1}{2 y^{2}} \left(- {}^{(1)}\C - {}^{(2)}\C + {}^{(3)}\C  + a_{1}^{2} y^{2} + a_{2}^{2} y^{2}  - a_{3}^{2} y^{2} \right. \nonumber \\
& \phantom{\frac{1}{2 y^{2}}}  + 2 a_{1} y  + 2 a_{2} y - 2 a_{3} y \Big).
\end{align}

\subsection{Simultaneous determination of the linear acceleration and the angular velocity of the frame}

\begin{figure}
\begin{center}
\includegraphics[scale = 0.4]{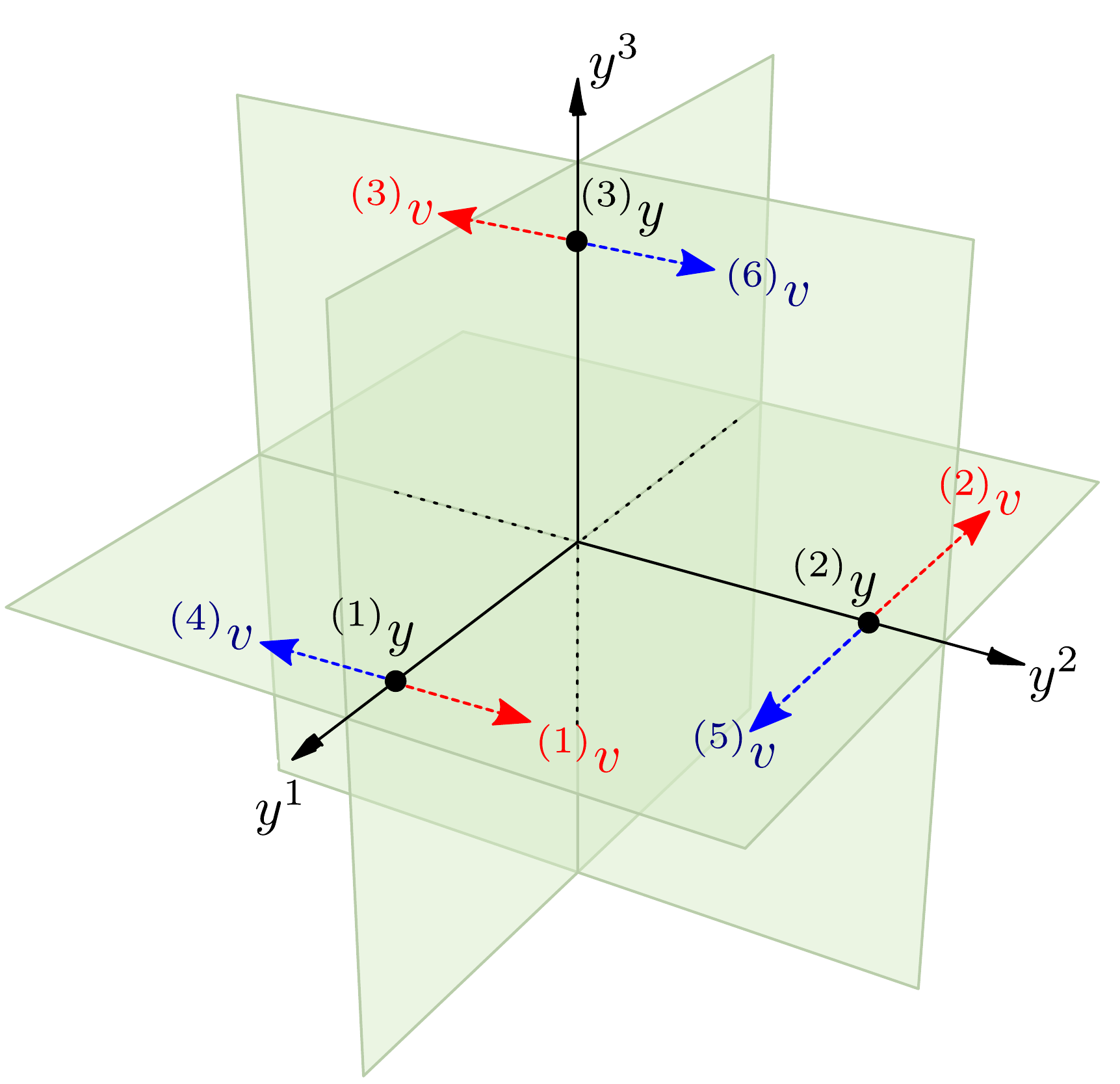}
\end{center}
\caption{\label{pos_3} Clock configuration used in the simultaneous determination of the linear acceleration \eqref{accel-both} and the angular velocity \eqref{omega-both}.}
\end{figure}

Here we extend the analysis from \cite{Puetzfeld:etal:2018:2} in order to find a configuration which allows to \textit{simultaneously obtain the linear acceleration and the angular velocity}, i.e.\ to compute all six components ($a_\alpha,\omega_\alpha$) using a suitable arrangement of clocks. This is achieved by considering pairs of clocks located along each axis and choosing velocities with opposite direction, perpendicular to their position vectors. Hence, the initial conditions for the 6 clocks are chosen as follows: the first 3 are located at ${}^{(1)}y^{\alpha}$, ${}^{(2)}y^{\alpha}$, and ${}^{(3)}y^{\alpha}$ as defined in \eqref{position_setup}, with velocities perpendicular to each position vector, i.e.\ we choose
\begin{eqnarray}
\nonumber {}^{(1)}v^{\alpha} = \left(\begin{array}{c} 0 \\ v_{12}\\ v_{13}\\ \end{array} \right), \quad {}^{(2)}v^{\alpha}=\left(\begin{array}{c} v_{21} \\ 0 \\ v_{23}\\ \end{array} \right),
\quad {}^{(3)}v^{\alpha} = \left(\begin{array}{c} v_{31} \\ v_{32}\\ 0\\ \end{array} \right). \label{rot_init2}
\end{eqnarray} 
The other 3 clocks are located at the same positions, but with velocities opposite to the first group. The situation is illustrated in Fig.\ \ref{pos_3}.

For this configuration of clocks, we obtain the following expressions for the angular velocity and the acceleration of the frame:
\begin{widetext}
\begin{align}
\nonumber \omega_{1} &= - \frac{v_{1 2} v_{3 1} y_{11} y_{33} \left(\pre{\C}{2,2} - \pre{\C}{2,-2} \right) + v_{1 3} v_{2 1} y_{11} y_{22}  \left(\pre{\C}{3,3} - \pre{\C}{3,-3} \right) + v_{2 1} v_{3 1} y_{22} y_{33} \left(\pre{\C}{1,1} - \pre{\C}{1,-1} \right)}{4 y_{11} y_{22} y_{33} \left(v_{1 2} v_{2 3} v_{3 1} - v_{1 3} v_{2 1} v_{3 2} \right)}, \\
 \omega_{2} &= - \frac{v_{1 2} v_{2 3} y_{11} y_{22} \left(\pre{\C}{3,3} - \pre{\C}{3,-3} \right) + v_{1 2} v_{3 2} y_{11} y_{33} \left(\pre{\C}{2,2} - \pre{\C}{2,-2} \right) + v_{2 1} v_{3 2} y_{22} y_{33} \left(\pre{\C}{1,1} - \pre{\C}{1,-1} \right)}{4 y_{11} y_{22} y_{33} \left(v_{1 2} v_{2 3} v_{3 1} - v_{1 3} v_{2 1} v_{3 2} \right)},  \label{omega-both} \\
\nonumber \omega_{3} &= - \frac{v_{1 3} v_{2 3} y_{11} y_{22}  \left(\pre{\C}{3,3} - \pre{\C}{3,-3} \right) + v_{1 3} v_{3 2} y_{11} y_{33} \left(\pre{\C}{2,2} - \pre{\C}{2,-2} \right) + v_{2 3} v_{3 1} y_{22} y_{33} \left(\pre{\C}{1,1} - \pre{\C}{1,-1} \right)}{4 y_{11} y_{22} y_{33} \left(v_{1 2} v_{2 3} v_{3 1} - v_{1 3} v_{2 1} v_{3 2} \right)},
\end{align}
and 
\begin{align}
 \nonumber a_1 &= \frac{1}{2y_{11}} \left(\sqrt{2  \pre{C}{1,1} + 2  \pre{C}{1,-1} + 4 \omega_{2}^{2} y_{11}^{2} + 4 \omega_{3}^{2} y_{11}^{2} + 4 v^{2}_{1} + 4} - 1\right), \\
 a_2 &= \frac{1}{2y_{22}} \left(\sqrt{2  \pre{C}{2,2} + 2  \pre{C}{2,-2} + 4 \omega_{1}^{2} y_{22}^{2} + 4 \omega_{3}^{2} y_{22}^{2} + 4 v^{2}_{2} + 4} - 1\right), \label{accel-both} \\
 \nonumber a_3 &= \frac{1}{2y_{33}} \left(\sqrt{2  \pre{C}{3,3} + 2  \pre{C}{3,-3} + 4 \omega_{1}^{2} y_{33}^{2} + 4 \omega_{2}^{2} y_{33}^{2} + 4 v^{2}_{3} + 4} - 1\right),
\end{align}
\end{widetext}
where one could insert \eqref{omega-both} into \eqref{accel-both} in order to obtain an explicit final result. Note that the minus sign ($-$) in the velocity indices indicates opposite velocity.

\section{Curvature determination}\label{sec_curvature_determination}

In order to find a similar analytical expression for the curvature components, we rearrange \eqref{general_freq_ratio_definition_our} as follows:
\begin{align}
\pre{y^\alpha}{n} \pre{y^\beta}{n} &  \left( - R_{0 \alpha \beta 0} - \frac{4}{3} \pre{v^\gamma}{m} R_{ \gamma \alpha \beta 0} \right.\nonumber\\
& \left.- \frac{1}{3} \pre{v^\gamma}{m} \pre{v^\delta}{m} R_{\alpha \gamma \delta \beta } \right) \nonumber \\
 &= B_3(\pre{y^\alpha}{n},\pre{v^\gamma}{m},a^\alpha, \omega^\alpha) = \pre{B_3}{n,m}, \label{curvature_master}
\end{align}
where we now define $B_3$ as
\begin{align}
 \pre{B_3}{n,m} &:= \pre{\C}{n,m} + \pre{v^2}{m} - 2 a_\alpha \pre{y^\alpha}{n} \nonumber \\
 &\quad - \pre{y^\alpha}{n} \pre{y^\beta}{n} \left( a_\alpha a_\beta - \delta_{\alpha \beta} \omega_\gamma \omega^\gamma + \omega_\alpha \omega_\beta \right) \nonumber \\
 &\quad - 2 v^\alpha \varepsilon_{\alpha \beta \gamma}  \pre{y^\beta}{n} \omega^\gamma. \label{form-of-B}
\end{align} 
Notice that the particular combination of curvature components in Eq.\,\eqref{curvature_master}, when written in terms of the newly defined $\pre{B_3}{n,m}$, is the same as in \cite{Puetzfeld:etal:2018:2}. The impact of our different choice for the state of motion of the reference clock, which is now at at rest in contrast to the original one in \cite{Puetzfeld:etal:2018:2}, is the explicit form of $\pre{B_3}{n,m}$, see \eqref{form-of-B}. Therefore, the results below are useful in both cases.

\subsection{Obtaining the components of the curvature}\label{sec:OCC}

We choose the same orientations and velocities for each clock as in section IV.D.\ of \cite{Puetzfeld:etal:2018:2}, but now we include their proper distances to the central clock explicitly. For the positions we choose ${}^{(n)}y^{\alpha}$ ($n=1,\dots,6$) in \eqref{position_setup}, with all distances set to equal values, i.e.\ $y_{11}=y_{22}=y_{33}=y_{41}=y_{42}=y_{52}=y_{52} = y_{61} = y_{63} = y$. Furthermore, we choose the following specific values for the velocities:
\begin{eqnarray}
&&{}^{(1)}v^{\alpha}=\left(\begin{array}{c} v_{11} \\ 0\\ 0\\ \end{array} \right), \qquad {}^{(2)}v^{\alpha}=\left(\begin{array}{c} 0 \\ v_{22}\\ 0\\ \end{array} \right), \\ 
&& {}^{(3)}v^{\alpha}=\left(\begin{array}{c} 0 \\ 0\\ v_{33}\\ \end{array} \right), \qquad {}^{(4)}v^{\alpha}=\left(\begin{array}{c} v_{41}\\ v_{42}\\ 0\\ \end{array} \right), \\ 
&& {}^{(5)}v^{\alpha}=\left(\begin{array}{c}  0\\ v_{52}\\ v_{53}\\ \end{array} \right), \qquad {}^{(6)}v^{\alpha}=\left(\begin{array}{c}  v_{61}\\ 0\\ v_{63}\\ \end{array} \right). \label{velocity_setup_curv}
\end{eqnarray}
In this solution, the components of the curvature are obtained by using clocks with positions and velocities which differ from the ones in \cite{Puetzfeld:etal:2018:2}. For each clock which is not at rest, its velocity is chosen perpendicular to its the position w.r.t.\ the central clock. This choice was motivated by the equivalence to the configuration in which each clock is (instantaneously) rotating around the reference clock. 

The first 6 components to be obtained are those corresponding to the constrained clock compass, i.e., the configuration of 6 clocks at rest discussed in section IV.F of \cite{Puetzfeld:etal:2018:2}:
\begin{align}
  R_{0110} &= - \frac{\pre{B_3}{1,0}}{y^{2}}, \label{curv_r0110} \\
  R_{0220} &= - \frac{\pre{B_3}{2,0}}{y^{2}}, \label{curv_r0220} \\
  R_{0330} &= - \frac{\pre{B_3}{3,0}}{y^{2}}, \label{curv_r0330} \\
  R_{0120} &= - \frac{1}{2 y^{2}} \left(\pre{B_3}{4,0} + y^2 ( R_{0110} + R_{0220} ) \right), \label{curv_r0120}\\
  R_{0130} &= - \frac{1}{2 y^{2}} \left(\pre{B_3}{6,0} + y^2 ( R_{0110} + R_{0330} ) \right), \label{curv_r0130}\\
  R_{0230} &= - \frac{1}{2 y^{2}} \left(\pre{B_3}{5,0} + y^2 ( R_{0220} + R_{0330} ) \right). \label{curv_r0230}
\end{align}
Here the index $0$ in the $\pre{B_3}{n,0}$ terms denotes clocks at rest (as before, the position and the velocity indices are also indicated). The 6 curvature components in the group above are those which can be determined using clocks at rest\footnote{The first 3 components can also be determined using clocks with velocities parallel to the respective position, then $R_{0110} = - \pre{B_3}{1,1}/y^{2}$, $R_{0220} = -\pre{B_3}{2,2}/y^{2}$ and $R_{0330} = -\pre{B_3}{3,3}/y^{2}$.}. 
The 14 remaining independent curvature components can be obtained as
\begin{align}
 R_{1210} &= \frac{3}{8 v_{2 2} y^{2}} \left(\pre{B_3}{1,2} - \pre{B_3}{1,-2}\right), \label{curv_r1210} \\ 
 R_{1310} &=  \frac{3}{8 v_{33} y^{2}}\left(\pre{B_3}{1,3} - \pre{B_3}{1,-3}\right), \label{curv_r1310} \\ 
 R_{2320} &= \frac{3}{8 v_{33} y^{2}} \left(\pre{B_3}{2,3} - \pre{B_3}{2,-3}\right), \label{curv_r2320} \\ 
 R_{1212} &= \frac{3}{2 v^2_{2 2}  y^{2} } \left(\pre{B_3}{1,2} + \pre{B_3}{1,-2} + 2 y^2 R_{0110}  \right), \label{curv_r1212} \\ 
  R_{1313} &=  \frac{3}{2 v^2_{33}  y^{2} } \left(\pre{B_3}{1,3} + \pre{B_3}{1,-3} + 2 y^2 R_{0110}  \right),  \label{curv_r1313}\\ 
  R_{2323} &= \frac{3}{2 v^2_{33}  y^{2} } \left(\pre{B_3}{2,3} + \pre{B_3}{2,-3} + 2 y^2 R_{0220}  \right), \label{curv_r2323}
\end{align}
\begin{widetext}
\begin{align}
 R_{1220} &= \frac{1}{4 v_{1 1} y^{2}}\left(- 3 \pre{B_3}{2,1} -3 R_{0220}y^{2} +  R_{1212}v_{1 1}^2 y^{2}\right), \label{curv_r1220}\\ 
 R_{1330} &= \frac{1}{4 v_{1 1} y^{2}}\left(- 3 \pre{B_3}{3,1} - 3 R_{0330}y^2 + R_{1313}v_{1 1}^2 y^{2}\right), \label{curv_r1330}\\
 R_{2330} &= \frac{1}{4 v_{2 2} y^{2}} \left(- 3\pre{B_3}{3,2}- 3 R_{0330}y^2 + R_{2323}v_{2 2}^2 y^{2}\right), \label{curv_r2330}\\
 R_{1213} &= \frac{1}{2 v_{5 2} v_{5 3} y^{2}} \left(3 \pre{B_3}{1,5} + 3 R_{0110} y^{2} - v_{5 2} y^{2} \left(4 R_{1210} + R_{1212} v_{5 2}\right) - v_{5 3} y^{2} \left(4 R_{1310} + R_{1313} v_{5 3}\right)\right), \label{curv_r1213}\\ 
 R_{1223} &= \frac{1}{2 v_{6 1} v_{6 3} y^{2}} \left(-3 \pre{B_3}{2,6} - 3 R_{0220} y^{2} + v_{6 1} y^{2} \left(R_{1212} v_{6 1} - 4 R_{1220}\right) + v_{6 3} y^{2} \left(4 R_{2320} + R_{2323} v_{6 3}\right)\right), \label{curv_r1223}\\ 
 R_{1323} &= \frac{1}{2 v_{4 1} v_{4 2} y^{2}} \left(3 \pre{B_3}{3,4} + 3 R_{0330} y^{2} + v_{4 1} y^{2} \left(- R_{1313} v_{4 1} + 4 R_{1330}\right) + v_{4 2} y^{2} \left(- R_{2323} v_{4 2} + 4 R_{2330}\right)\right), \label{curv_r1323}\\
  R_{1230} &= \frac{1}{4 v_{3 3} y^{2}}\left(-3 \pre{B_3}{4,3} - 3(R_{0110}+2R_{0120}+ R_{0220})y^2 + 4(R_{1310}+R_{2320})v_{3 3} y^{2} \right. \nonumber\\
  &\qquad\qquad\qquad + (R_{1313}+ 2R_{1323} + R_{2323})v_{3 3}^2 y^{2}\Big) , \label{curv_r1230}\\
R_{2310} &=  \frac{1}{4 v_{11} y^{2}}\left(3 \pre{B_3}{5,1} + 3(R_{0220}+2R_{0230}+ R_{0330})y^2 + 4(R_{1220}+R_{1330})v_{11} y^{2} \right.\nonumber\\
  &\qquad\qquad\qquad - (R_{1212}+ 2R_{1213} + R_{1313})v_{11}^2 y^{2}\Big) . \label{curv_r2310} 
\end{align}
\end{widetext}

Again, the minus sign ($-$) in the velocity index indicates opposite velocity. This allows us to determine the 20 independent components of the curvature by means of 20 different clocks/measurements. Note that this solution is different from the one presented in \cite{Puetzfeld:etal:2018:2}.

The solution in \eqref{curv_r0110}-\eqref{curv_r2310} relates measurements that need to be performed and the physical parameters, i.e. the curvature components, in a hierarchical way. This means that we write the expression of some curvature components in terms of previously determined ones, plus the outcome of new clock measurements. As will be discussed later, one can use this hierarchy as a possible strategy to determine the different curvature components. Alternatively, one can obtained ``direct'' expressions for each curvature component in terms of the measurements, by replacing the corresponding previous components. The result is displayed in Eqs.\ \eqref{R0110B}--\eqref{R2310B} in Appendix \ref{RB}.

The solution \eqref{curv_r0110} tells us that the component $R_{0110}$ can be determined by means of one clock at rest located along the $x$-axis, so that both values $\pre{B_3}{1,0}$ and $y$ have to be known. We denote this clock configuration by $(1,0)$. The components $R_{0220}$ and $R_{0330}$ can be determined analogously, this time by means of the configurations $(2,0)$ and $(3,0)$, which are located along the $y$- and the $z$-axis, respectively. 

In order to determine $R_{0120}$ one needs measurements from more than one clock. As is apparent from Eq.\ \eqref{curv_r0120}, in addition to the knowledge of $R_{0110}$ and $R_{0220}$, one needs data from  measurements with a clock at rest located in the $xy$-plane at a 45 degree angle from the $x$ and $y$ axis (which we denote by $(4,0)$) -- c.f.\ also ${}^{(4)}y^{\alpha}$ in \eqref{position_setup}, with $y_{41}=y_{42}=y$. Equivalently, $R_{0120}$ can be directly determined with the frequency data from three clocks: $(1,0)$, $(2,0)$ and $(4,0)$, see Eq.\ \eqref{R0120B}. Notice that in this case an additional ``simultaneous'' determination of $R_{0110}$, $R_{0220}$ and $R_{0120}$ is also possible, starting from the data of the same configurations, i.e.\ $(1,0)$, $(2,0)$ and $(4,0)$. The determination of $R_{0130}$ and $R_{0230}$ can be performed in an analogous fashion, by defining a second group, see Eqs.\,\eqref{curv_r0130} and \eqref{curv_r0230} respectively and/or \eqref{R0130B} and \eqref{R0230B}.  Notice that, as can be seen from Eq.\ \eqref{general_freq_ratio_definition_our},  in a more general situation, when considering clocks with generic positions in the $x$--$y$ plane, we will need measurements of clocks with at least 3 different positions in order to decouple the contribution of the components $R_{0110}$, $R_{0220}$ and $R_{0120}$ from the quantity $\C$. 

A third group of measurements is defined by \eqref{curv_r1210}--\eqref{curv_r2320}. Each of these components can be computed from data of two different clock configurations: in the case of $R_{1210}$ by measurements of clocks in configurations $(1,2)$ and $(1,-2)$, and similarly for $R_{1310}$ and $R_{2320}$.

A fourth group is given by \eqref{curv_r1212}--\eqref{curv_r2323}. These expressions show that, for instance, $R_{1212}$ can be determined with data of $R_{0110}$ and the result of measurements of the clocks configurations $(1,2)$ and $(1,-2)$. Equivalently, the direct determination needs 3 clock configurations: $(1,0)$, $(1,2)$ and $(1,-2)$; i.e.\ from a combination of the data from the first and third group above, see Eq.\ \eqref{R1212B}. This also means that the group of configurations $(1,0)$, $(1,2)$ and $(1,-2)$ suffices to ``simultaneously'' determine the three components $R_{0110}$, $R_{1210}$ and $R_{1212}$. A similar relation holds for the group $R_{0110}$, $R_{1310}$ and $R_{1313}$ and the configurations $(1,0)$, $(1,3)$ and $(1,-3)$, as well as for $R_{0220}$, $R_{2320}$ and $R_{2323}$ and the configurations $(2,0)$, $(2,3)$ and $(2,-3)$, see Eqs.\ \eqref{curv_r1310} and \eqref{curv_r1313}, as well as \eqref{curv_r2320} and \eqref{curv_r2323}, respectively. 

A fifth group of curvature components, represented by \eqref{curv_r1220}--\eqref{curv_r2330}, can be obtained by using two previously determined curvature components, plus data from one new clock. The alternative direct determination, as shown in Eqs.\ \eqref{R1220B}--\eqref{R2330B}, requires a total of five clock configurations. Alternatively, one may perform a simultaneous determination of $R_{0110}$, $R_{0220}$, $R_{1210}$, $R_{1212}$ and $R_{1220}$ with the help of the five configurations $(1,0)$, $(1,2)$, $(1,-2)$, $(2,0)$ and $(2,1)$. Such simultaneous measurements are also possible for the group $R_{0110}$, $R_{0330}$, $R_{1310}$, $R_{1313}$ and $R_{1330}$, by using the configurations $(1,0)$, $(1,3)$, $(1,-3)$, $(3,0)$ and $(3,1)$; as well as for the group $R_{0330}$, $R_{0220}$, $R_{2320}$, $R_{2323}$ and $R_{2330}$, by utilizing $(2,0)$, $(2,3)$, $(2,-3)$, $(3,0)$ and $(3,2)$.

A sixth group is given by \eqref{curv_r1213}--\eqref{curv_r1323}, in which the curvature can be obtained from previous data plus data from one additional clock -- $R_{1213}$ requires measurements from the $(1,5)$, $R_{1223}$ from the $(2,6)$, and $R_{1323}$ from the $(3,4)$ configuration. The fully resolved ``direct'' expressions are shown in Eqs.\ \eqref{R1213B}--\eqref{R1323B}. As with the previous groups, one could also perform a simultaneous determination of the curvature from the measurements of a group of suitably chosen clock configurations. As an example, we infer that the configurations $(1,0)$, $(1,2)$, $(1,-2)$, $(1,3)$, $(1,-3)$ and $(1,5)$ simultaneously determine $R_{1213}$, $R_{0110}$, $R_{1210}$, $R_{1212}$, $R_{1310}$ and $R_{1313}$. 

Finally, a seventh group is given by \eqref{curv_r1230} and \eqref{curv_r2310}. The determination of $R_{1230}$ and $R_{2310}$ requires only measurements from one additional clock, in addition to the previous configurations. As an example, the determination of $R_{1230}$ requires data from the $(4,3)$ configuration. Again, the fully replaced expressions for those components can be found in appendix \ref{RB}.

In table \ref{tab_number_measurements} the 20 curvature components are grouped by the structure of the solution, and by the number of required measurements. The choice of a hierarchical or simultaneous determination is going to play an important role in the error analysis, which we discuss in the next section.

\begin{table}
\caption{\label{tab_number_measurements}Number of measurements required for different curvature components.}
\begin{ruledtabular}
\begin{tabular}{cccc}
Curvature components & Group & \# Measurements\\
\hline
 $R_{0110}$,  $R_{0220}$,  $R_{0330}$& 1 & 1\\
 $R_{0120}$,  $R_{0130}$,  $R_{0230}$& 2 & 3\\
 $R_{1210}$,  $R_{1310}$,  $R_{2320}$& 3 & 2\\
 $R_{1212}$,  $R_{1313}$,  $R_{2323}$& 4 & 3\\
 $R_{1220}$,  $R_{1330}$,  $R_{2330}$& 5 & 5\\
 $R_{1213}$ & 6 & 6\\
 $R_{1223}$ & 6 & 8\\
 $R_{1323}$ & 6 & 10\\
 $R_{1230}$, $R_{2310}$ & 7 & 12\\
\end{tabular}
\end{ruledtabular}
\end{table}

\section{Simulated parameter estimation}\label{types_of_measurements_sec}

In this section we perform simulations in order to illustrate how the different parameters could be determined. Additionally we estimate the precision with which we can measure each physical quantity. In particular, we show how parameter changes impact the determination of the acceleration, the angular velocity, and the curvature, by using simulated data.

\subsection{Data generation}\label{sec_data_gen}

In order to perform a simulation, we need to create data for each clock (position and velocity), and a model for the measurable variable, i.e.\ the proper time ratio, which in turn determines the value of $\bar{C}$. 

For the mock data set we generate, for each clock configuration, $N$ values for its position and velocity, assuming a normal distribution for both variables ($y \sim {\mathcal N} (\bar{y}, \sigma_y^2 )$, $v \sim {\mathcal N} (\bar{v}, \sigma_v^2 )$). With these values, and the assumed test values for the quantities which we want to determine (acceleration, angular velocity and curvature components), we obtain, by means of the master equation \eqref{general_freq_ratio_definition_our}, the corresponding values of the frequency ratio $\bar{C}$ for each of the $N$ clocks. In a subsequent step we add noise to the $\bar{C}$ values, thereby modeling the uncertainty of the measurement process of the proper times. For the noise we also assume a Gaussian distribution $\delta \bar{C} \sim {\mathcal N} (0, \sigma^2_{\bar{C}})$ with vanishing mean. The standard deviation is set to the intrinsic instability of the clock, as reported for example in \cite{Chou:etal:2010:1}. Furthermore, we assume that the errors in the frequency ratio, position and velocity are independent of each other.

\subsection{Determination of the linear acceleration}\label{sec_a}

First we perform a simulation in order to show how the acceleration of the reference system could be determined. We consider a reference frame moving with constant (time-independent) acceleration $a^\alpha$ in the direction of the $x$-axis, assuming a test value $a_1=-9.8\ \rm m/s^2$. For the angular velocity we assume the same orientation as the acceleration, so that $\omega_1 = 7.3 \times 10^{-5}\ \rm rad/s$ (which would correspond to the local angular velocity due to Earth's rotation at the north pole). 

Taking into account the experimental results for state of the art clocks given in \cite{Chou:etal:2010:2} and \cite{Chou:etal:2010:1}, we work out the errors for the frequency ratio variable $\C$ when measured by such clocks. Chou et al.\ \cite{Chou:etal:2010:2} reported a fractional frequency inaccuracy of $8.6\times 10^{-18}$ for optical clocks, and gave in \cite{Chou:etal:2010:1} an error of $1.6\times 10^{-17}$ for the fractional frequency change of optical clocks with difference in height of 33 cm, due to relativistic effects. Note that, due to \eqref{meaning_of_c_approx}, the absolute error of the proper time ratio variable $\C$ is twice the value of the absolute error of the redshift. In the following, we use that value as the standard deviation of our assumed normal distribution, $\sigma_{\C} = 3.2\times 10^{-17}$, in our simulations.

Considering the above, we perform a simulation using a mock data set, generated as explained in section \ref{sec_data_gen}, using an array of clocks at rest w.r.t.\ the reference clock -- which we previously worked out in section \ref{paragraph_fermi_normal_flat_linear_accel_determination}. We  consider $N = 100$ samples of measurements (for each pair of clocks, see section \ref{paragraph_fermi_normal_flat_linear_accel_determination}), with mean distances $\bar{y}$ ranging from $0.37\,\rm m$ (the distance reported in \cite{Chou:etal:2010:1}) to $10.5\,\rm m$, and with $\sigma_y=1\, \rm cm$. For simplicity we consider a vanishing mean value and standard deviation for the velocity (i.e.\ $\bar{v}=\sigma_v=0$). Following \cite{Chou:etal:2010:1} we set $\sigma_{\C} = 3.2\times 10^{-17}$. Using the data generated in this way we then determine the probability distribution for the acceleration $a_1$, by using a Markov chain Monte Carlo (MCMC) method, as implemented in the \texttt{EMCEE} Python package \cite{emcee:2013}. For the inference of $a_1$ we use a Gaussian likelihood together with flat priors in combination with the master equation \eqref{general_freq_ratio_definition_our}.  

\begin{figure}
\begin{center}
\includegraphics[scale = 0.6]{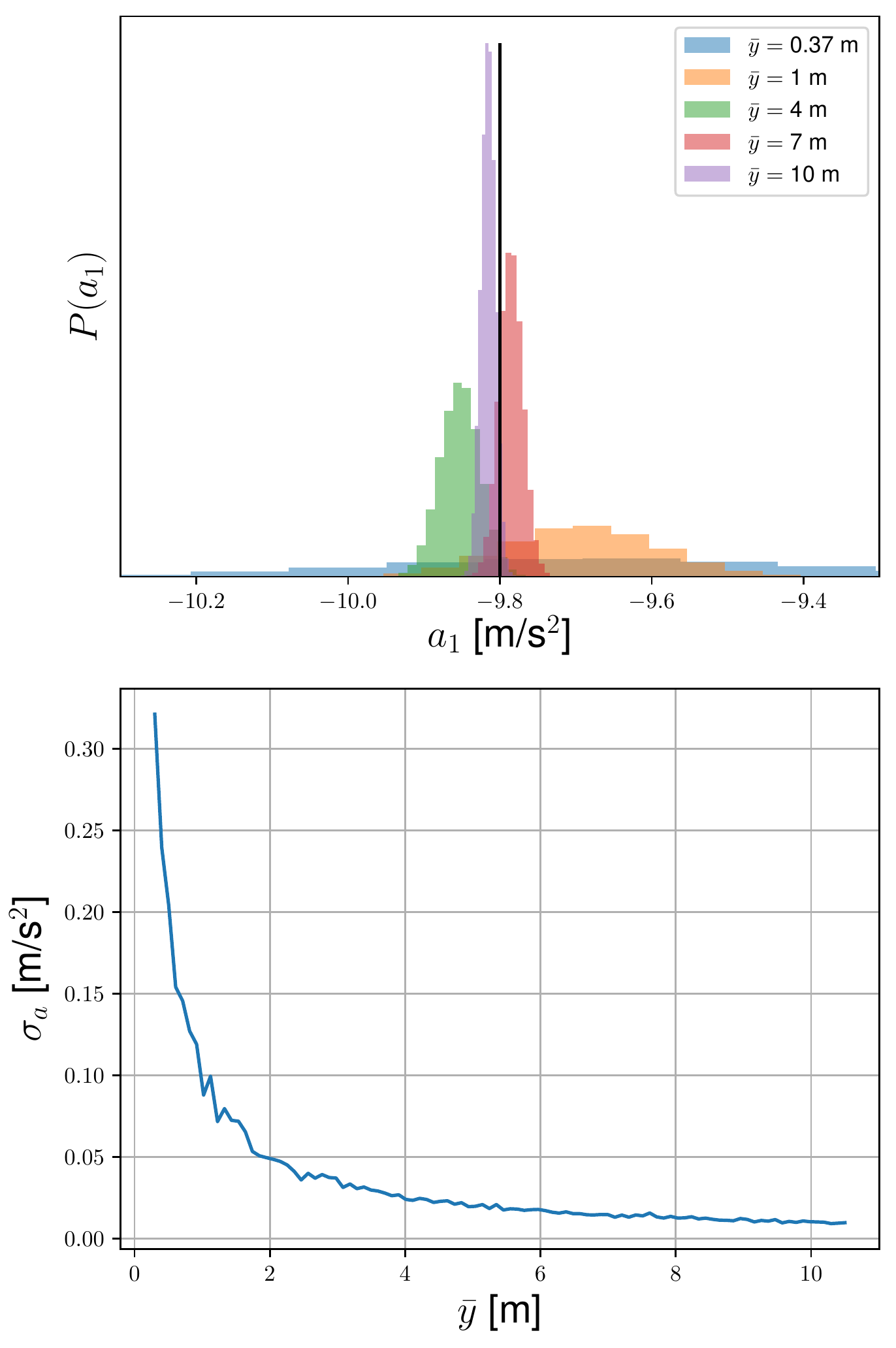}
\end{center}
\caption{\label{fig_4} Posterior for $a_1$ for different representative values of the mean clock height. For this calculation, we use $\sigma_{\C} = 3.2\times 10^{-17} $, and $\sigma_y = 1 {\rm\,cm}$. The black vertical line represents the test value of $a_1 = -9.8 {\rm m/s}^2$, $N = 100$.}
\end{figure}

A representative example of the posterior for $a_1$ for a set of simulated measurements, and for different values of $\bar{y}$ is shown in Fig.\,\ref{fig_4}. As expected, with increasing separation of the clocks the variance of the inferred values of $a_1$ decreases, and the mean value approaches the assumed test value. An increment of the mean distance from $\bar{y}=1\,\rm m$ to $\bar{y}=10\,\rm m$ from the reference clock reduces the standard deviation of the acceleration from $\pm 0.10 {\rm\, m/s^2}$ to $\pm 0.011 {\rm\, m/s^2}$. 

\begin{figure}
\begin{center}
\includegraphics[scale = 0.6]{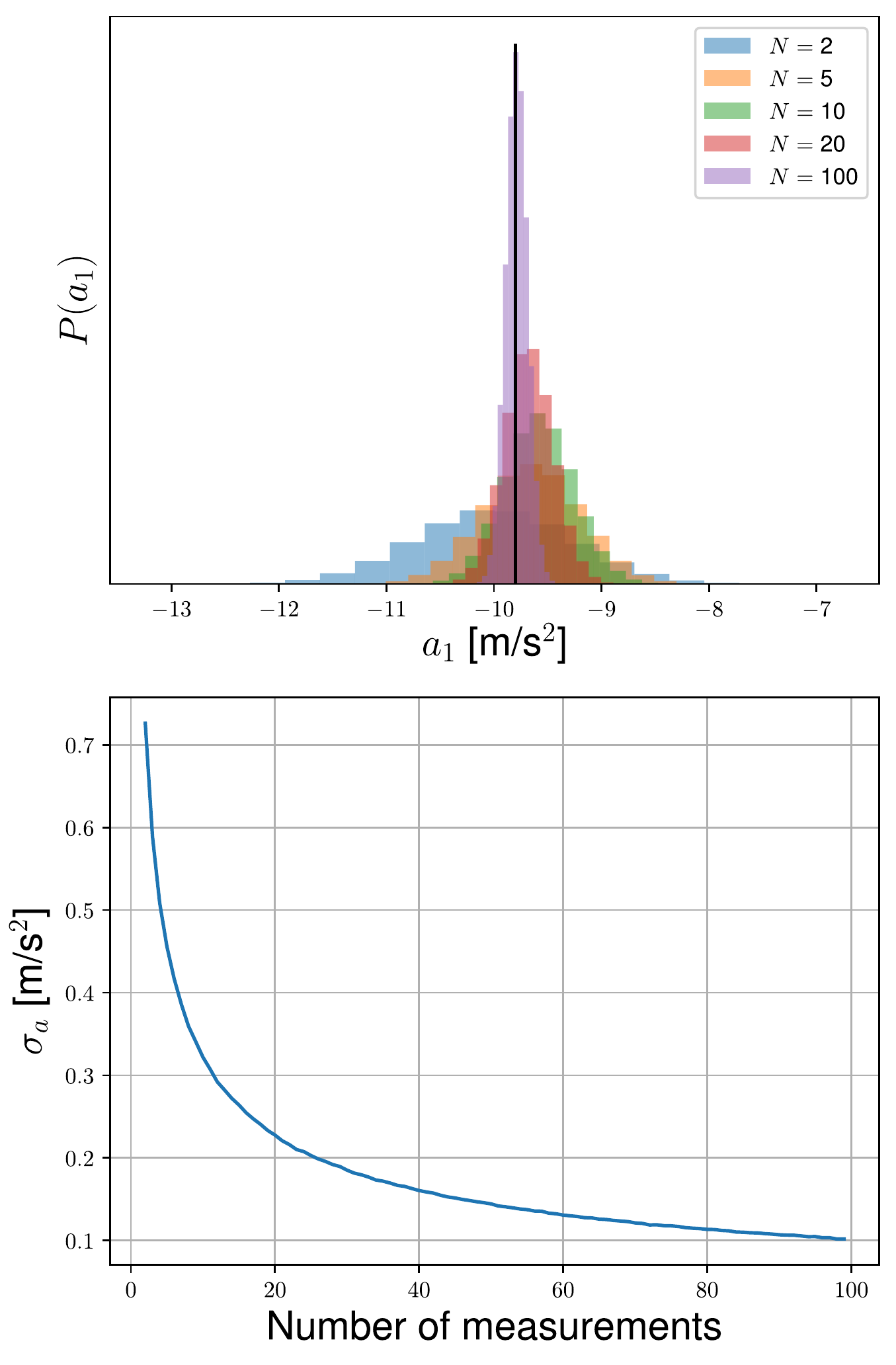}
\end{center}
\caption{\label{fig5accel} Posterior for $a_1$ for different representative values of the number of initial samples, from 2 to 100. For this calculation, we use $\sigma_{\C} = 3.2\times 10^{-17} $, $\bar{y} = 1 {\rm\,m}$ and $\sigma_y = 1 {\rm\,cm}$. The black vertical line represents the assumed test value of $a_1 = -9.8 {\rm m/s}^2$.}
\end{figure}

Additionally, we perform the calculation varying the number $N$ of clock measurements, with the same initial conditions as in the previous case, but now setting $\bar{y} = 1 {\rm\,m}$. The result is shown in Fig.\ \ref{fig5accel}. As expected, the precision in the determination of $a_1$ increases with the number of measurements. For instance, with $N=100$ measurements, we obtain a value of $\sigma_a$ of the order of $0.1\, {\rm m/s}^2$.

\subsection{Determination of the curvature components}\label{error-curv}

In this section we present results which illustrate the measurement strategy and the precision with which curvature components could be obtained. As a simplification, we will suppose that our reference system is free falling, i.e., $a^\alpha = 0$ and $\omega^\alpha = 0$.

We will use the Schwarzschild metric as a guide for the computation of the mock values of the curvature. The non-vanishing components for the Riemann curvature tensor of Schwarzschild spacetime in Schwarzschild coordinates $(ct,r,\theta,\varphi)$ are 
\begin{align}
 R_{trrt} &= \dfrac{r_{\rm s}}{r^3}, \\
 R_{\theta \varphi \theta \varphi } &= r r_{\rm s} \sin^2 \theta, \\
 2 R_{r \theta r \theta} &= \dfrac{r_{\rm s}}{r_{\rm s} - r}, \\
 2 R_{r \varphi r \varphi} &= \dfrac{r_{\rm s}}{r_{\rm s} - r} \sin^2 \theta, \\
 2 R_{t \theta \theta t} &= -\dfrac{r_{\rm s}(r - r_{\rm s})}{r^2}, \\
 2 R_{t \varphi \varphi t} &= -\dfrac{r_{\rm s}(r - r_{\rm s})}{r^2} \sin^2 \theta,
\end{align}
where $r_s=2GM/c^2$ is the Schwarzschild radius. Considering an orthonormal basis whose spacelike vectors $e_1$, $e_2$ and $e_3$ are aligned along the $r$, $\theta$ and $\varphi$ directions respectively. We then obtain (for further details see, for instance, Ref.\ \cite{Bini:2005})
\begin{align}
R_{0110} = R_{2323} &= \dfrac{r_{\rm s}}{r^3}, \label{curvs01}\\
R_{0220} = R_{0330} = R_{1212} = R_{1313} &= - \dfrac{r_{\rm s}}{2r^3}. \label{curvs02}
\end{align}
In our simulations we would like to consider the component $R_{0120}$ as non-vanishing in order to deal with non-zero numerical quantities in our subsequent examples. Therefore, we assign the value $R_{0120}=R_{0110}/3$ for this component by hand. In summary, in our simulations we shall use the following non-vanishing test values:
\begin{align}
R_{0110} &=  3.415 \times 10^{-23}\ \text{m}^{-2}, \label{R0110test} \\
R_{0220} &=  -1.708 \times 10^{-23}\ \text{m}^{-2}, \label{R0220test} \\
R_{0120} &=  1.138 \times 10^{-23}\ \text{m}^{-2}. \label{R0120test}
\end{align}
Notice that we choose $r$ equal to the radius of the Earth, so that the curvature components are of the order of the curvature produced by our planet on its surface.

\subsubsection{Obtaining one curvature component} \label{error-curv-real}

\begin{figure}
\begin{center}
\includegraphics[scale=0.9]{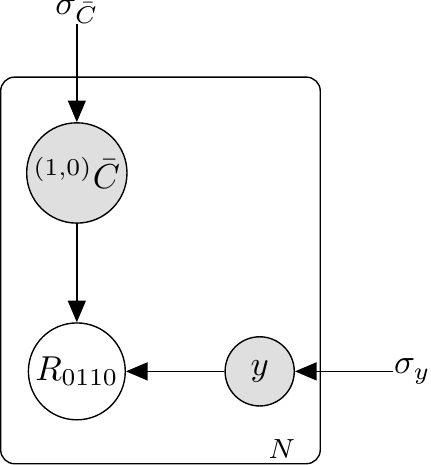}
\caption{First type of measurement: $R_{0110}$. Similarly for $R_{0220}$ and $R_{0330}$.}
\label{fig:Diag1}
\end{center}
\end{figure}

We start with the simplest case in which we can determine a single component of the Riemannian curvature tensor, using only one clock configuration, as in the first group discussed in section \ref{sec_curvature_determination}, for instance $R_{0110}$. In this case that component is determined by the value and uncertainty of the distance $y$, as well as the auxiliary quantity $\C$. The statistical structure of this first type of measurement is illustrated in Fig.\ \ref{fig:Diag1}.

\begin{figure}
\begin{center}
\includegraphics[scale = 0.5]{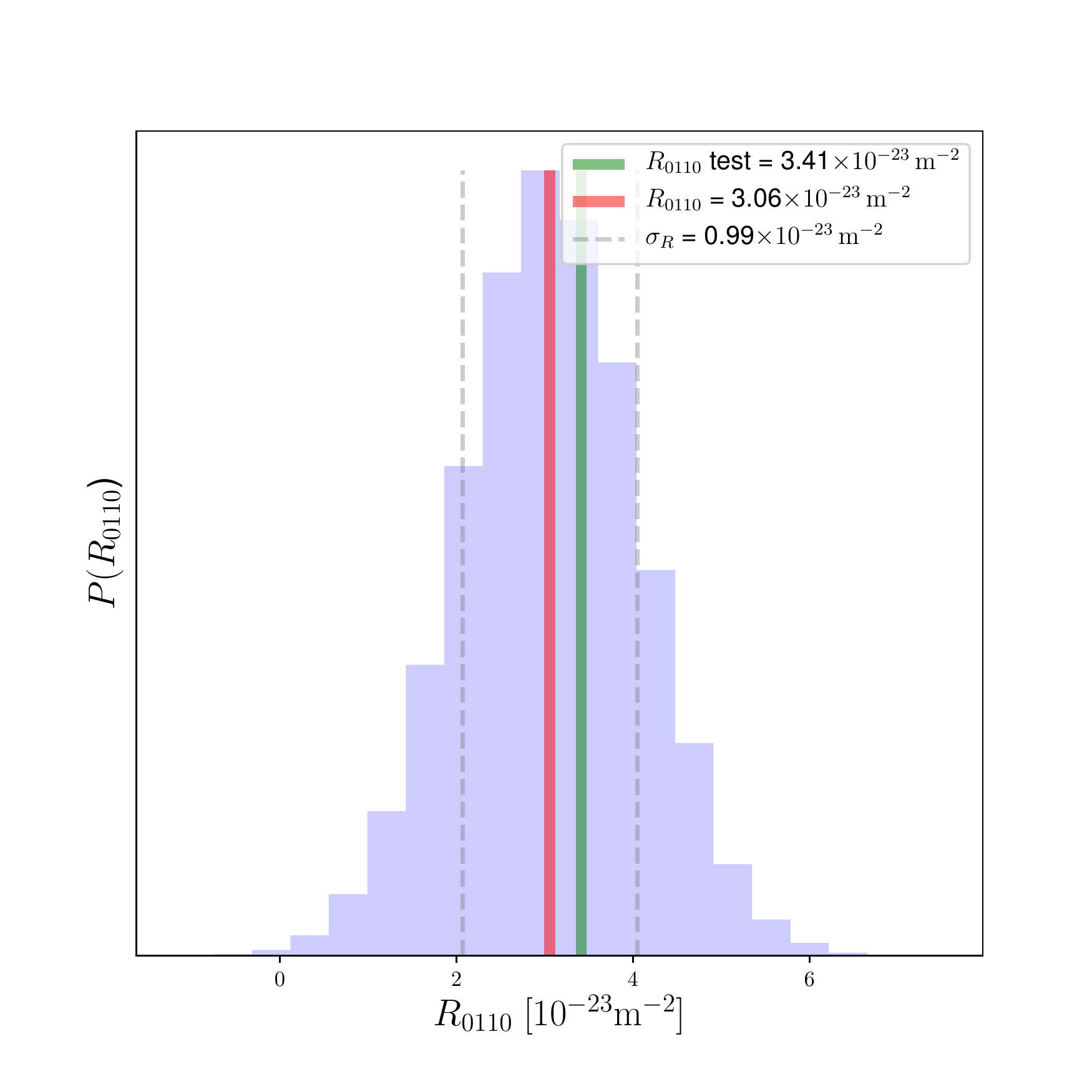}
\end{center}
\caption{\label{Fig6} Posterior for the curvature component $R_{0110}$ using realistic parameter values (see text). The red vertical line represents the mean of the distribution. The grey vertical dashed lines represent the percentiles 16 and 84, which for a Gaussian distribution corresponds to the interval $[{\rm mean} \pm 1\sigma]$. The green vertical line represents the test value \eqref{R0110test}. We consider $N = 100$, with $y \sim {\mathcal N} (\bar{y} = 10\, {\rm km}, \sigma^2_{y} = 10^4 \, {\rm m}^2)$ and $\sigma_{\bar{C}} = 10^{-14}$.}
\end{figure}

We now perform a simulation with $N = 100$ simulated proper time ratio measurements, with positions $y \sim {\mathcal N} (\bar{y} = 10\,{\rm km}, \sigma^2_{y} = 10^4 \,\rm m^2)$; as well as a normally distributed $\bar{C}$, with $\mu_{\C}$ given by the master equation, and $\sigma_{\bar{C}} = 10^{-14}$, which could be considered as a moderately optimistic value since it is three orders of magnitude higher than the precision reported by \citep{Chou:etal:2010:2} for experiments with fairly ideal and controlled conditions. The results for representative mock data are shown in Fig.\ \ref{Fig6}. The distribution of values for the curvature are characterized by a mean value of $\bar{R}_{0110} = 3.06\times 10^{-23}\,\text{m}^{-2}$, and a standard deviation of $\sigma_{R} = 0.99\times 10^{-23}\,\text{m}^{-2}$.  

\subsubsection{Varying parameters}

\begin{figure}
\begin{center}
\includegraphics[scale = 0.6]{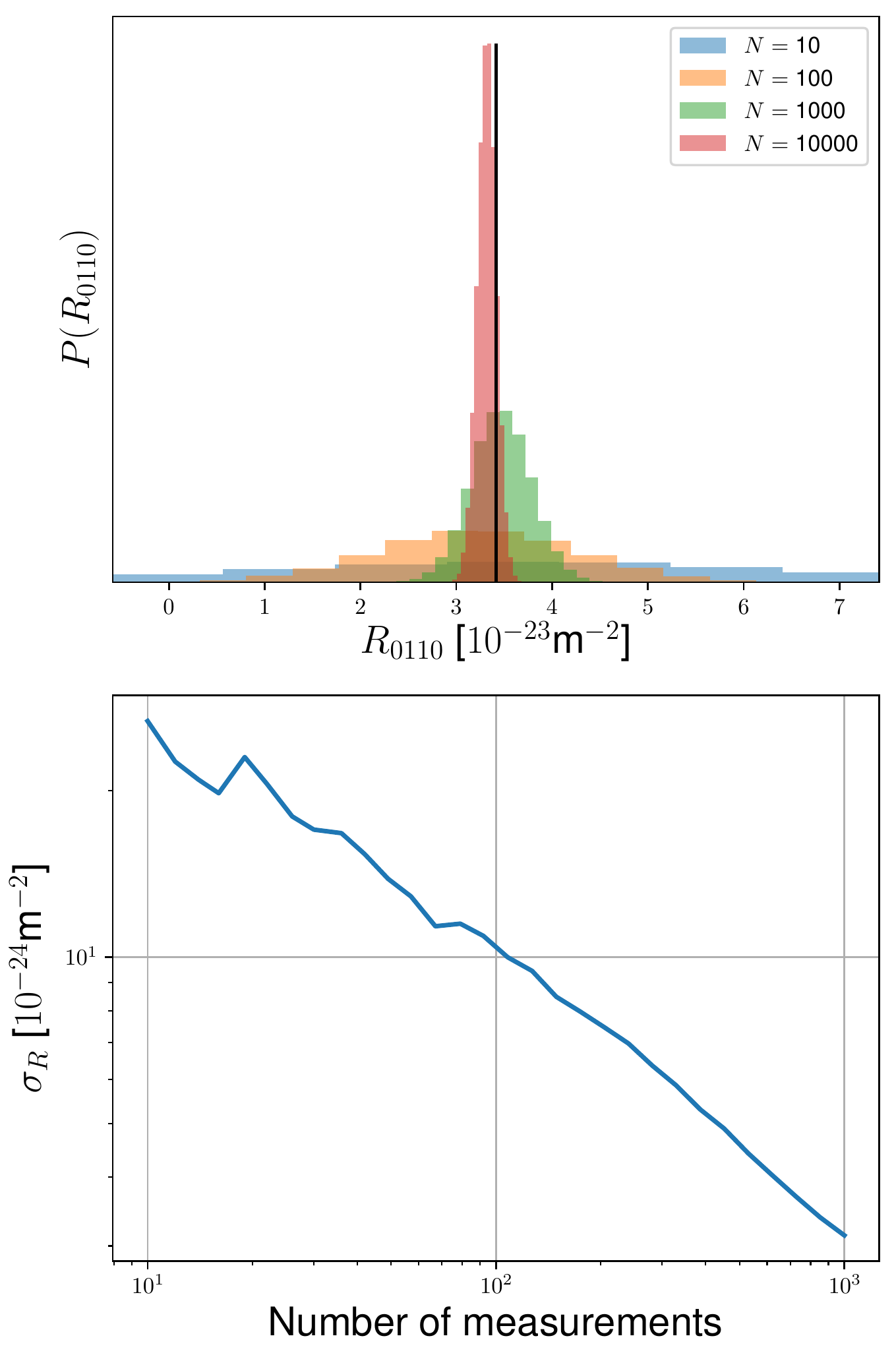}
\end{center}
\caption{\label{Fig7} The upper plot contains the variation of the probability distribution for the curvature component $R_{0110}$ for the indicated number of measurements. We see how the mean value for the curvature component approaches the test value \eqref{R0110test}. In the lower plot (log-log scale) the standard deviation is shown to decrease with increasing number of measurements. We have used $\bar{y}=10\, {\rm km}$, $\sigma_y=100\, {\rm m}$, and $\sigma_{\C} = 10^{-14}$ as input parameters.}
\end{figure}

\begin{figure}
\begin{center}
\includegraphics[scale = 0.6]{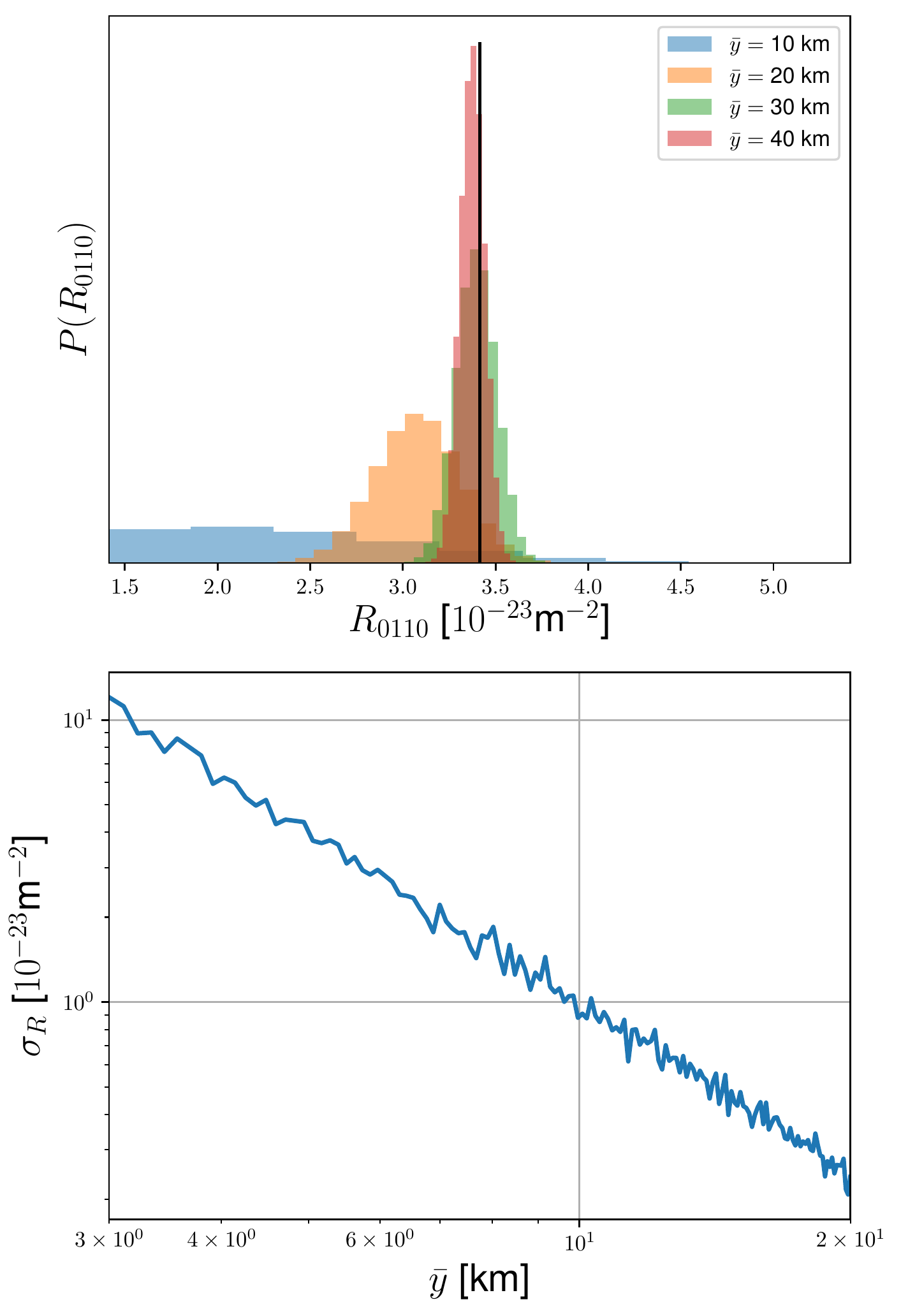}
\end{center}
\caption{\label{Fig8} The upper plot contains the variation of the probability distribution for the curvature component $R_{0110}$ for the indicated values of the distance/position of the clock. The lower plot (log-log scale) shows the standard deviation of the distribution for different values of $\bar{y}$. We used $N=100$, $\sigma_y=100\,\rm m$, and $\sigma_{\bar C}=10^{-14}$.}
\end{figure}

\begin{figure}
\begin{center}
\includegraphics[scale = 0.6]{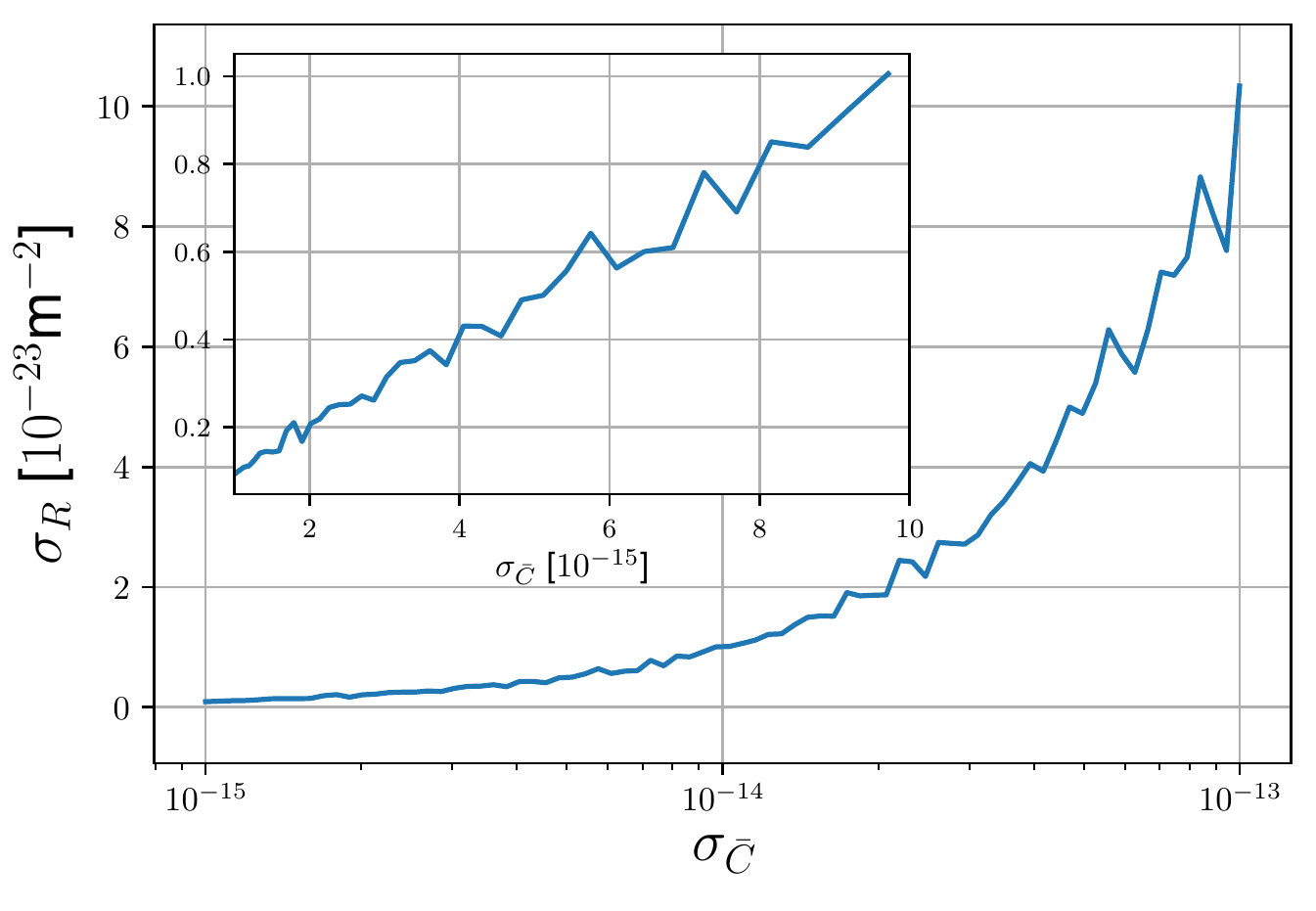}
\end{center}
\caption{\label{Fig11} Standard deviation of the probability distribution of the curvature component $R_{0110}$ for different standard deviations of the measurement of the frequency ratio $\bar{C}$. The outer plot shows the evolution in the interval of $\sigma_{\C}$ from $10^{-15}$ to $10^{-13}$, in logarithmic scale for the $x$-axis. The inner plot shows the evolution of the same variable over the interval of $\sigma_{\C}$ from $10^{-15}$ to $10^{-14}$, on a linear scale. We have used $N = 100$, $\bar{y}  = 10\,{\rm km}$, and $\sigma_y = 100\,\rm m$. }
\end{figure}

Here we determine how the probability distributions for the curvature component $R_{0110}$ change when varying some of the parameters of our simulation. Some representative results are shown in figures \ref{Fig7} to \ref{Fig11}. We can see from Figs.\ \ref{Fig7} and \ref{Fig8} that the precision of the curvature determination increases, as expected, with growing number of measurements as well as with the distance of the clocks to the origin (reference clock), while the mean of the distribution fluctuates as it approaches the assumed test value. For instance, for the input values used in our simulations we observed that the standard deviation of the posterior distribution for the curvature component decreases from $\approx 3 \times 10^{-23} \text{m}^{-2}$ for $N=10$ to $\approx 1 \times 10^{-23} \text{m}^{-2}$ for $N=100$, and finally to $\approx 3 \times 10^{-24} \text{m}^{-2}$ for $N=1000$. These values are consistent with a decay of the expected form $\sigma_{R} \sim {N}^{-1/2}$. Similarly we observe from Fig.\ \ref{Fig8} how $ \sigma_{R}$ decreases as $\bar{y}$ increases. For instance, for $\bar{y} = 10\text{\,km}$, $20\text{\,km}$, and $40\text{\,km}$, we obtain $ \sigma_{R} \approx 9\times 10^{-24}  \text{\,m}^{-2}$,  $2\times 10^{-24} \text{\,m}^{-2}$ and $7\times 10^{-25} \text{\,m}^{-2}$, respectively. This is consistent with the expected behavior of $\sigma_{R} \sim \bar{y}^{-2}$, see Eq.\ \eqref{apendix_C_1}.  

\subsubsection{Multiparameter Bayesian analysis}

\begin{figure}
\begin{center}
\includegraphics[scale=0.9]{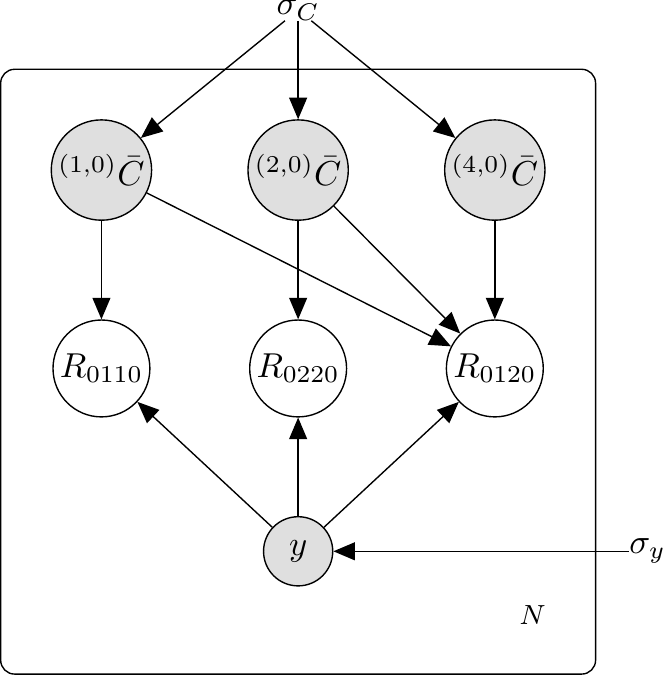}
\caption{Scheme for the simultaneous determination of $R_{0110}$, $R_{0220}$ and $R_{0120}$. The lines represent the dependencies given by Eqs.\ \eqref{R0110B}, \eqref{R0220B}, and \eqref{R0120B}.}
\label{fig:Diag2}
\end{center}
\end{figure}

Now, we determine one of the curvature components of the second group, see table \ref{tab_number_measurements}, for instance $R_{0120}$. In this case, we need three clocks with three different positions. Taking this into account, we simulate the simultaneous determination of $R_{0110}$, $R_{0220}$, and $R_{0120}$. We obtain distributions for these three curvature components, from simulated measurements of the proper time ratios of clocks at rest, with positions as discussed in section \ref{sec:OCC}. The determination of the curvature components are affected by the value of the distance $y$ and of the auxiliary quantities ${}^{(1,0)}\C$, ${}^{(2,0)}\C$ and ${}^{(4,0)}\C$, and their uncertainties. The general dependency of this second kind of curvature determination is depicted in Fig.\ \ref{fig:Diag2}.

\begin{figure}
\begin{center}
\includegraphics[scale = 0.45]{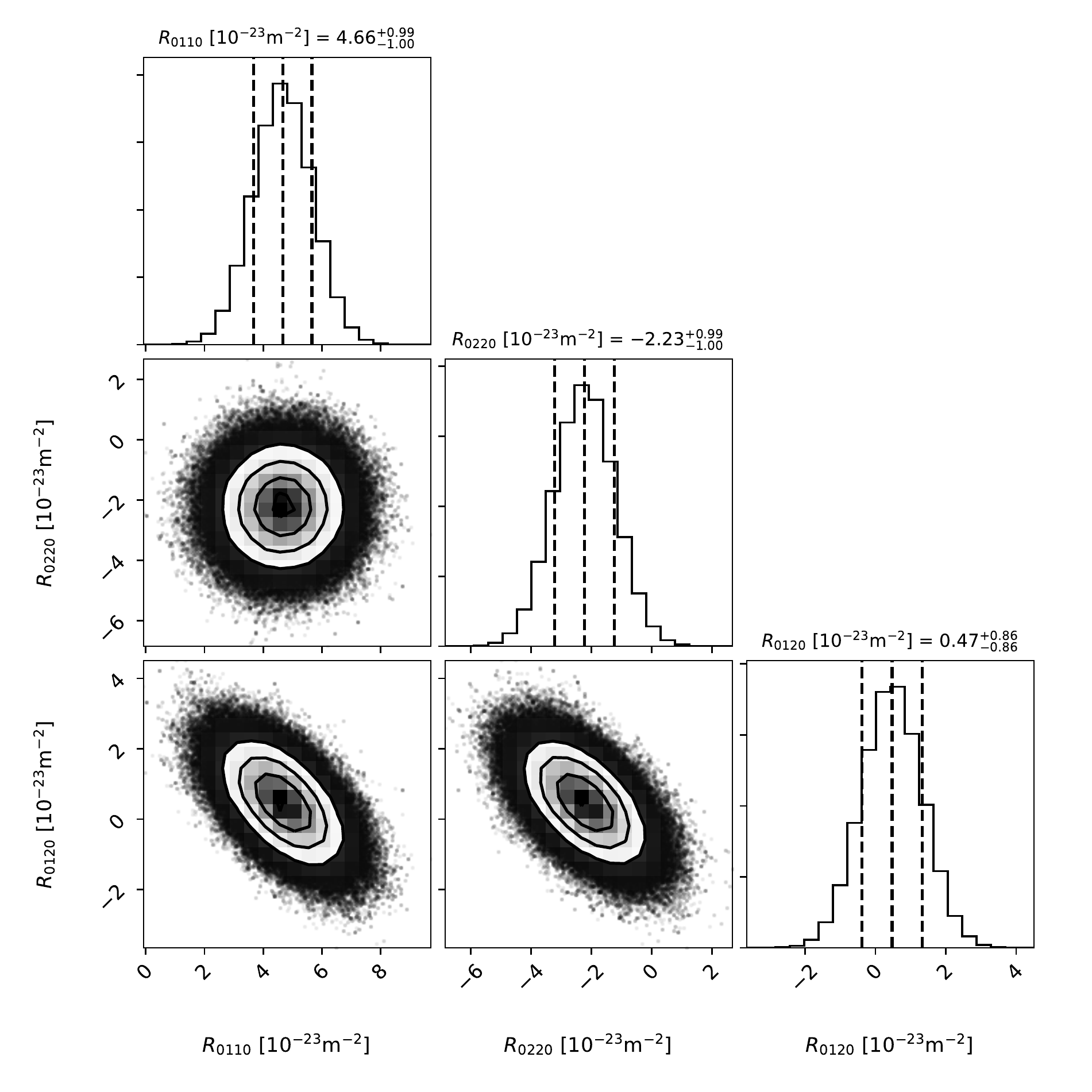}
\end{center}
\caption{\label{fig_sim_0110-0220-0120-N10} Probability distribution for the curvature components $R_{0110}$, $R_{0220}$ and $R_{0120}$, obtained simultaneously. We used $\bar{y} = 10\,\rm km$, $\sigma_{y_x} = \sigma_{y_y} = 100\,\rm m$, $\sigma_{\bar{C}} = 10^{-14}$, and $N = 100$ measurements for each clock in configurations $(1,0)$, $(2,0)$ and $(4,0)$  (i.e.\ $3\times 100$ clocks, positions, and frequency ratio values).}
\end{figure}

\begin{figure}
\begin{center}
\includegraphics[scale = 0.6]{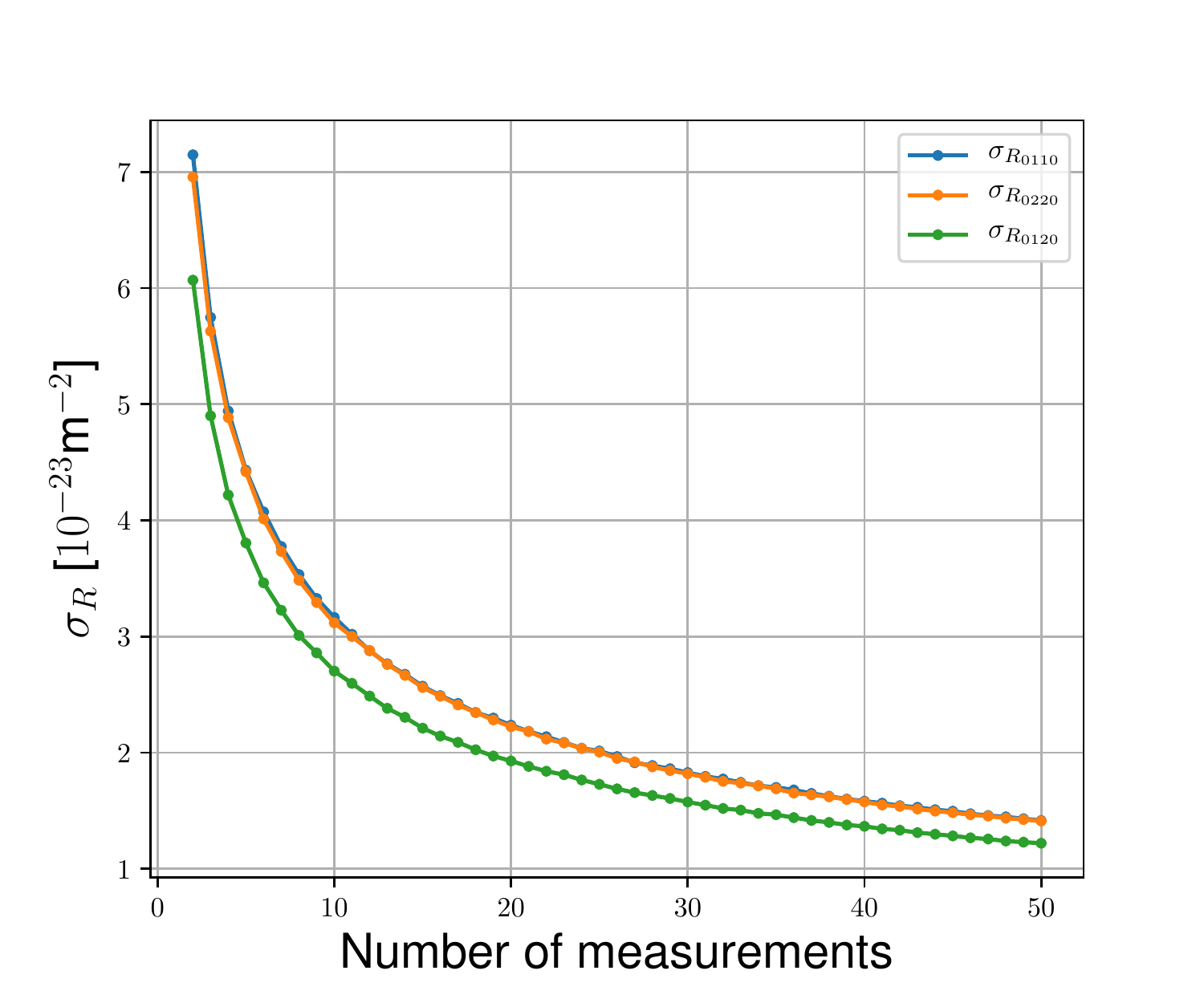}
\caption{\label{sim-loop-N-R0120} Evolution of the standard deviation for the simultaneous determination of the curvature components $R_{0110}$, $R_{0220}$ and $R_{0120}$ for different number of measurements. We used $\bar{y} =10\,\rm km$, $\sigma_{y_x} = \sigma_{y_y}= 100\,\rm m$, $\sigma_{\bar{C}} = 10^{-14}$.}
\label{Fig_stds_R0110-R0220-R0120_varying_N}
\end{center}
\end{figure}

We simulate the values of the curvature using expressions \eqref{R0110test}--\eqref{R0120test}, while the other components are set to zero. We also use the same parameters for the clocks as in section \ref{error-curv-real}, which are, $N = 100$ (for each arrangement of 3 clocks) and $\bar{y} = 10\,\rm km$. The standard deviations are set to $\sigma_{y} = 100\,\rm m$ and $\sigma_{\bar{C}} = 10^{-14}$. Here we neglect the influence of the velocity of the clocks. The results are shown in Fig.\ \ref{fig_sim_0110-0220-0120-N10}, from which we can infer the standard deviation of the resulting distributions is the order of $10^{-23}\,\text{m}^{-2}$ for each component, and that the test input values lie within a $2\sigma$ interval. Fig.\ \ref{Fig_stds_R0110-R0220-R0120_varying_N} shows how the standard deviations of each of these three curvature components decreases when a higher number $N$ of measurements are used for the inference.

\begin{figure}
\begin{center}
\includegraphics[scale=0.8]{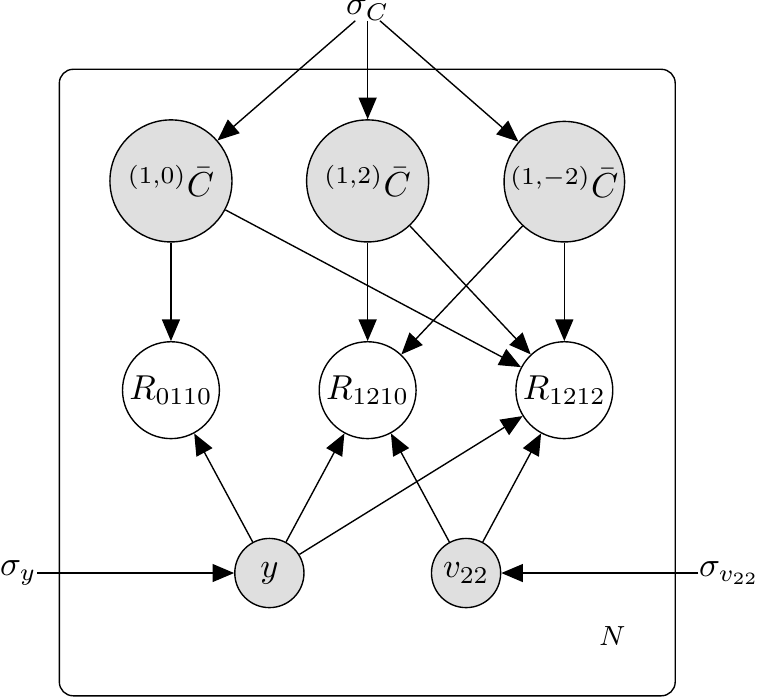}
\caption{Scheme for the determination of the group ($R_{0110}$, $R_{1210}$, $R_{1212}$) from the measurements of configurations $(1,0)$, $(1,2)$ and $(1,-2)$. The lines represent the dependency given by Eqs. \eqref{R0110B}, \eqref{R1210B} and \eqref{R1212B}. See Fig.\ \ref{fig_sim_0110-1210-1212-N10} for the results of the corresponding simultaneous determination.}
\label{fig:R0110-R1210-R1212}
\end{center}
\end{figure}

\begin{figure}
\begin{center}
\includegraphics[scale = 0.45]{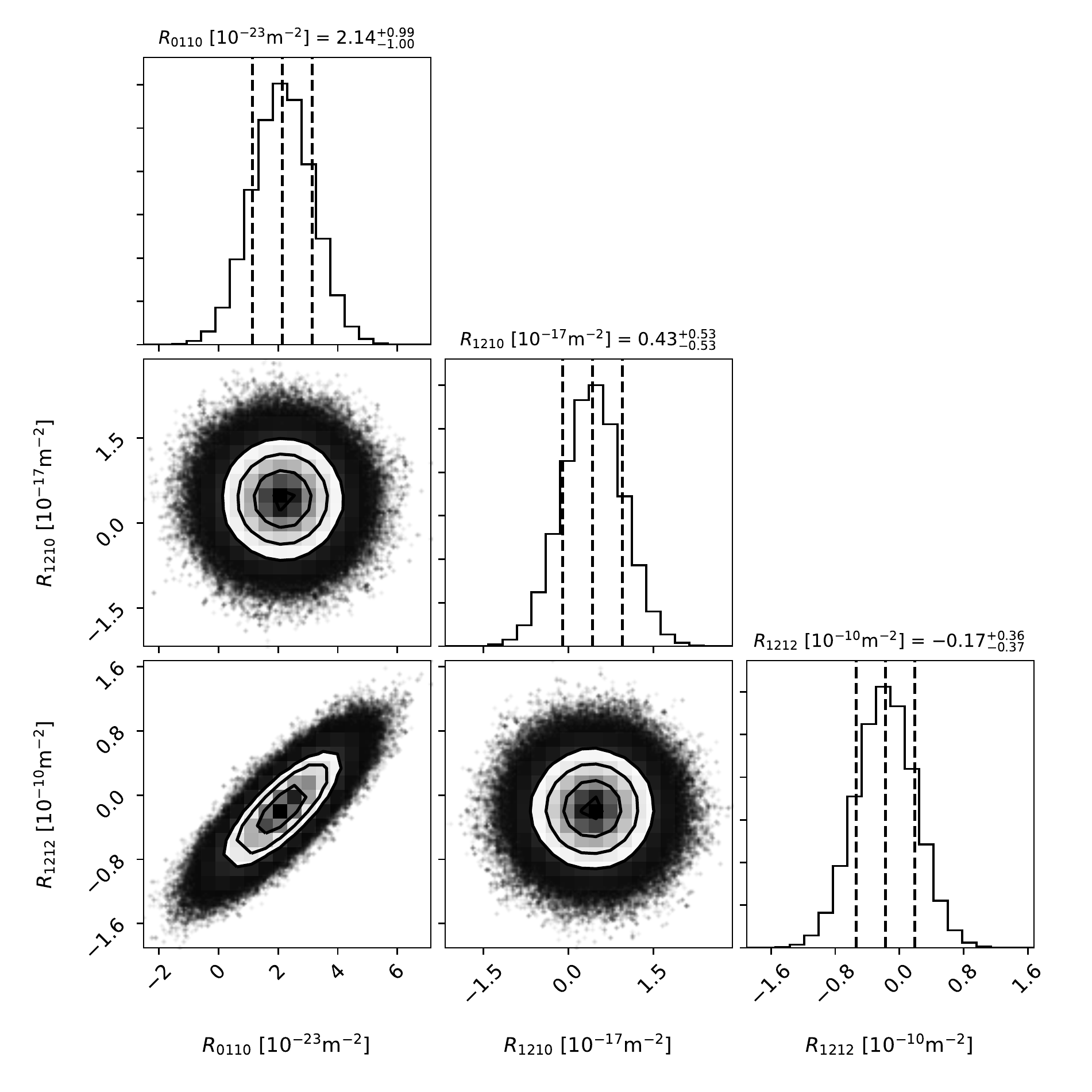}
\end{center}
\caption{\label{fig_sim_0110-1210-1212-N10} Probability distribution for the curvature components $R_{0110}$, $R_{1210}$ and $R_{1212}$, obtained simultaneously. We used $\bar{y} = 10\,\rm km$, $\sigma_{y} = 100\,\rm m$, $\bar{v}= 10^{-6}c$, $\sigma_v = 10^{-8}c$, $\sigma_{\bar{C}} = 10^{-14}$, and $N = 100$ measurements for each clock in configurations $(1,0)$, $(1,2)$ and $(1,-2)$ (i.e. $3\times 100$ clocks, positions, velocities, and frequency ratio values).}
\end{figure}

In Fig.\ \ref{fig:R0110-R1210-R1212} we show an example of the simultaneous inference of three curvature components, namely $R_{0110}$, $R_{1210}$ and $R_{1212}$, starting from data of 3 clock configurations. This case is qualitatively different from the first one since now two of the clocks (those corresponding to configurations $(1,2)$ and $(1,-2)$) are necessarily moving w.r.t.\ the central clock, which allows to infer values for $R_{1210}$ and $R_{1212}$, as discussed in detail in Section \ref{sec:OCC}. For this simulation we use $N = 100$ (for each arrangement of three clocks), $\bar{y} = 10\,\rm km$, $\sigma_{y} = 100 \, \rm m$, $\bar{v} = 10^{-6}c$, $\sigma_{v} = 10^{-8}c$, and $\sigma_{\bar{C}} = 10^{-14}$. The results are shown in Fig.\ \ref{fig_sim_0110-1210-1212-N10}. We observe that the test values \eqref{R0110test}--\eqref{R0120test} are indeed recovered within the corresponding $2\sigma$ intervals. Additionally, each curvature component is determined with a different precision: the standard deviation of the distribution for $R_{0110}$, $R_{1210}$ and $R_{1212}$ are of the order of $10^{-23}\text{\,m}^{-2}$, $10^{-17}\text{\,m}^{-2}$ and $10^{-11}\text{\,m}^{-2}$, respectively. This is a consequence of the additional effect of the velocity involved in the analysis, which reduces the precision of the determination of the curvature components ``with more spatial indices'', in a hierarchical way. This can be understood by looking at the master equation \eqref{general_freq_ratio_definition_our}, where the curvature component $R_{1210}$ contributes to the measurable frequency ratio with a term which is suppressed by a factor linear in the velocity $v/c\sim 10^{-6}$ when compared to $R_{0110}$, while the component $R_{1212}$ is suppressed by a term quadratic in $v/c$. 

\begin{figure}
\begin{center}
\includegraphics[scale=0.7]{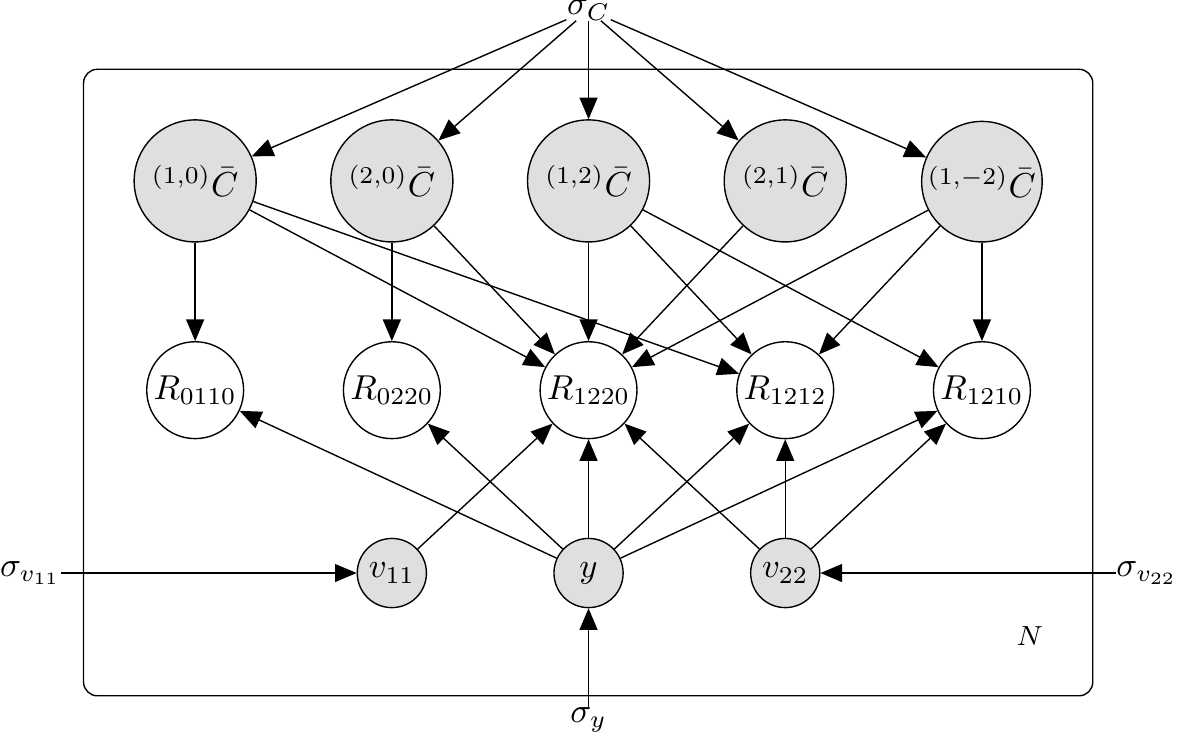}
\caption{Scheme for the determination of the group ($R_{0110}$, $R_{0220}$, $R_{1210}$, $R_{1220}$, $R_{1212}$) from the measurements of configurations $(1,0)$, $(2,0)$, $(1,2)$, $(2,1)$ and $(1,-2)$. The lines represent the dependencies given by Eqs.\ \eqref{R0110B}, \eqref{R0220B}, \eqref{R1210B}, \eqref{R1220B} and \eqref{R1212B}. See Fig.\ \ref{fig_sim_0110-0220-1210-1212-1220-N10} for the results of the corresponding simultaneous determination.}
\label{fig:5Rs}
\end{center}
\end{figure}

\begin{figure}
\begin{center}
\includegraphics[scale = 0.6]{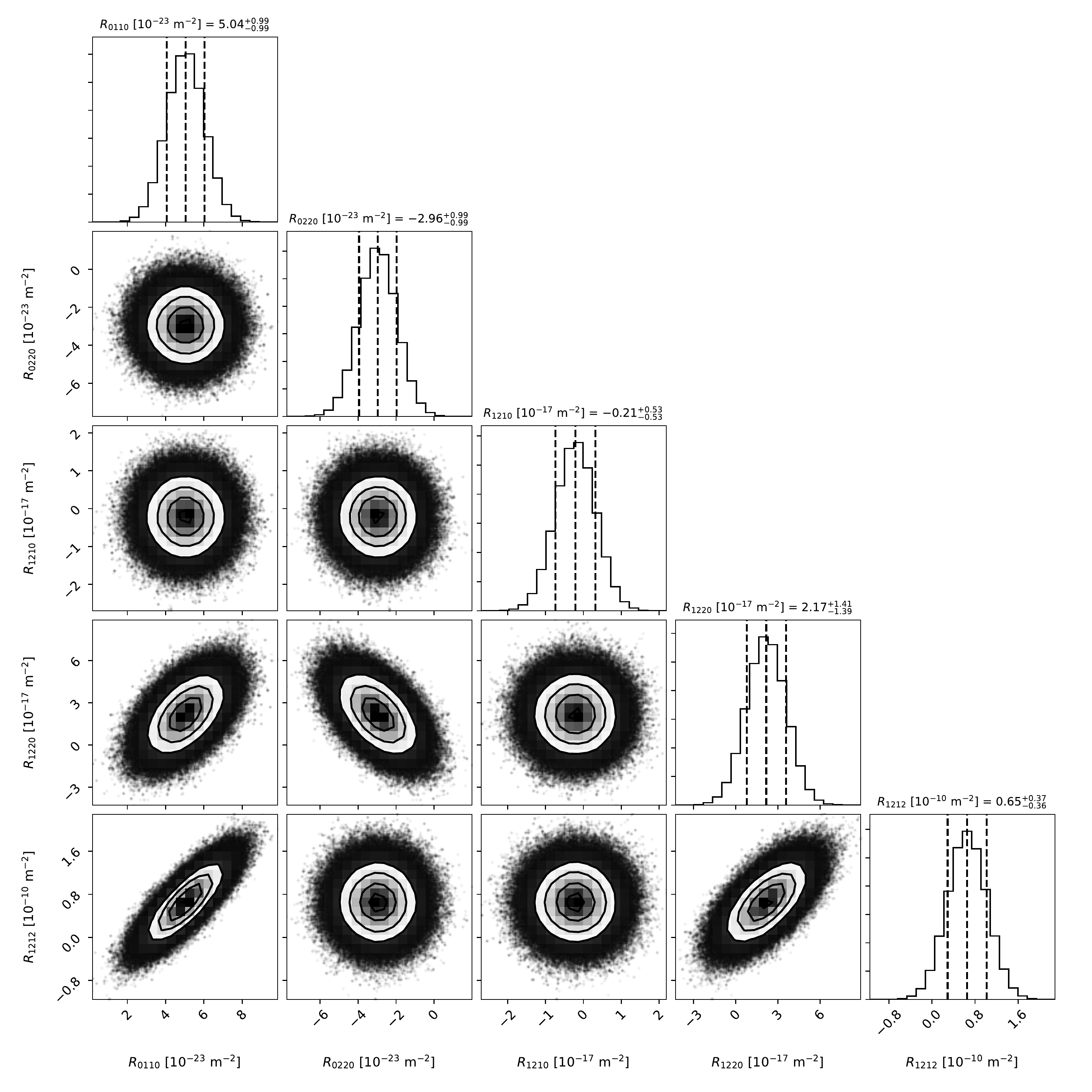}
\caption{\label{fig_sim_0110-0220-1210-1212-1220-N10} Obtained distribution for the curvature components $R_{0110}$, $R_{0220}$, $R_{1210}$, $R_{1212}$ and $R_{1220}$, obtained simultaneously. In this case, we used $\bar{y} =10\,\rm km$. $\sigma_y = 100\,\rm m$, $\sigma_{\bar{C}} = 10^{-14}$, $\bar{v} = 10^{-6}c$, and $\sigma_{v} = 10^{-8}c$ and $N = 100$ measurements for each clock in configurations $(1,0)$, $(2,0)$, $(1,2)$, $(1,-2)$ and $(2,1)$ (i.e.\ $5\times 100$ clocks, positions, velocities, and frequency ratio values).}
\end{center}
\end{figure}

We also show the results of a simultaneous inference of five curvature components, $R_{0110}$, $R_{0220}$, $R_{1210}$, $R_{1212}$, and $R_{1220}$, by using data from 5 clocks. See the discussion in section \ref{sec_curvature_determination} and  Fig.\ \ref{fig:5Rs} which illustrates the process. Using again $N = 100$ (for each arrangement of five clocks), $\bar{y} = 10\text{\,km}$, $\sigma_{y} = 100\,\rm m$, $\bar{v} = 10^{-6}c$, $\sigma_{v} = 10^{-8}c$, and $\sigma_{\bar{C}} = 10^{-14}$, the obtained result is shown in Fig.\ \ref{fig_sim_0110-0220-1210-1212-1220-N10}. The behavior of this more complex case is similar to the previous one, in the sense that the velocity defines a hierarchy of precisions for the determination of each curvature component: the distributions of $R_{0110}$ and $R_{0220}$ have a standard deviation, for the values used in our example, of the order of $10^{-23}\text{\,m}^{-2}$, while for $R_{1210}$ and $R_{1220}$ we obtain values of the order $10^{-17}\text{\,m}^{-2}$, and finally $R_{1212}$ is the component with the largest error of the order $10^{-10}\text{\,m}^{-2}$. If we compare how the component $R_{0110}$ is determined in this simultaneous determination with the result of the simpler cases with three curvatures (Fig.\ \ref{fig_sim_0110-1210-1212-N10}) and also with the single determination (Fig.\ \ref{Fig6}), we notice similar results for the spread of the corresponding obtained distribution. 

\begin{figure}
\begin{center}
\includegraphics[scale = 0.6]{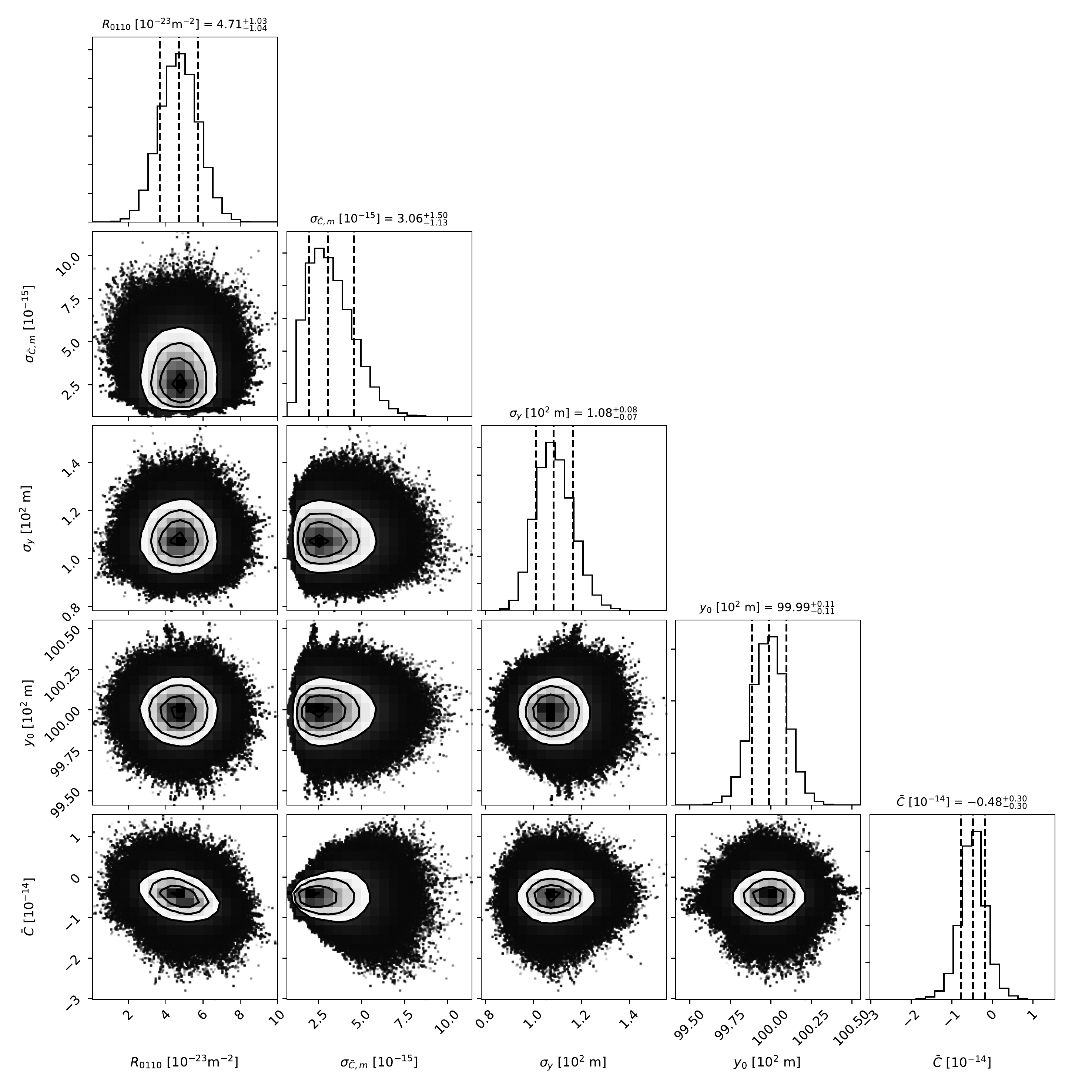}
\caption{\label{fig_withNoise_R0110_measurements} Posterior for the curvature component $R_{0110}$ and additional nuisance parameters, using realistic parameter values. We used $\bar{y} =10\,\rm km$, $\sigma_y = 100\,\rm m$, $\sigma_{\bar{C}} = 10^{-14}$, and $N=100$ measurements for each clock in configuration $(1,0)$ (i.e.\ $1\times 100$ clocks, positions, velocities, and frequency ratio values).}
\end{center}
\end{figure}

As an additional test we have also considered the addition of nuisance parameters in the posterior analysis. In Fig.\ \ref{fig_withNoise_R0110_measurements} we show the result for the simpler curvature component, $R_{0110}$, when we also fit the parameter $y_0$ and $\sigma_y$ as the mean value and standard deviation of the assumed Gaussian distribution of the distance $y$; $\C$ as the mean value of the Gaussian distribution of the frequency ratio variable $\C$ and $\sigma_{\C,m}$ as the standard deviation of the variation of the frequency ratio w.r.t.\ our model, given by the master Eq.\ \eqref{general_freq_ratio_definition_our}. The obtained values of these nuisance parameters are within the expected range, as well as the value for the component $R_{0110}$, and with a precision similar to that found in the simpler analysis in Fig.\ \ref{Fig6}, as well as to the result of the simultaneous curvature determination in Fig.\ \ref{fig_sim_0110-0220-1210-1212-1220-N10}. A similar result can be obtained for the $R_{0220}$ component.

\begin{figure}
\begin{center}
\includegraphics[scale = 0.5]{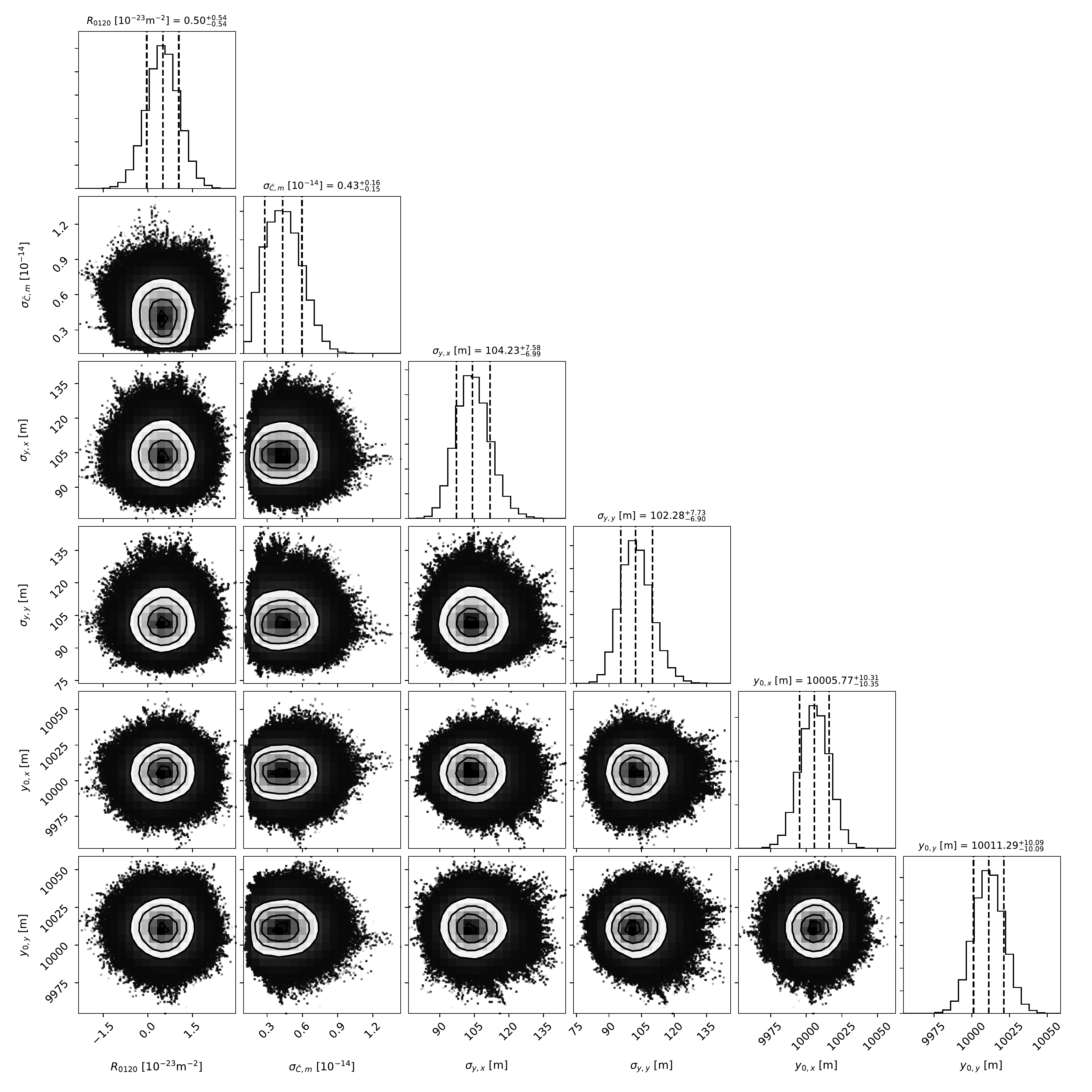}
\caption{\label{fig_withNoise_hierarchical-N200-R0120} Estimation of $R_{0120}$ in a hierarchical way, with nuisance parameters estimation. We used $\bar{y} = 10\,\rm km$, $\sigma_{y_x} = \sigma_{y_y}= 100\,\rm m$, $\sigma_{\bar{C}} = 10^{-14}$, and $N = 100$ (i.e.\ $1\times 100$ additional clocks, positions, and frequency ratio values in the $(4,0)$ configuration).}
\end{center}
\end{figure}

Finally, we estimate again the component $R_{0120}$ but, in contrast to the simultaneous determination shown in Fig.\ (\ref{fig_sim_0110-0220-0120-N10}), now we perform a hierarchical calculation with additional nuisance parameters. This means that we infer the distribution for $R_{0120}$ using the resulting distributions of the components $R_{0110}$ and $R_{0220}$, which were calculated as discussed in the paragraph above, as inputs. In Fig.\ (\ref{fig_withNoise_hierarchical-N200-R0120}) we show the result for this curvature component, when we also fit the parameters $y_{0,x}$ and $y_{0,y}$ as the mean values for the $x$ and $y$ position coordinates of the clock, respectively; $\sigma_{y,x}$ and $\sigma_{y,y}$ as the corresponding standard deviations; and $\sigma_{\C,m}$ as the standard deviation of the variation of the frequency ratio w.r.t.\ our model. The obtained values of these nuisance parameters are again within the expected range, as well as the value for the component $R_{0120}$. Looking at the standard deviation of the resulting probability distribution for $R_{0120}$ we obtain a value of the order of $10^{-23}\text{ m}^{-2}$, which is of the same order as the value resulting from the simultaneous determination presented above and displayed in Fig.\ (\ref{fig_sim_0110-0220-0120-N10}).

\section{Conclusions \& outlook}\label{sec_conclusions}

We have worked out a new solution, as well as a complete statistical description of the gravitational clock compass \cite{Szekeres:1965,Puetzfeld:Obukhov:2016:1}. The model of the compass presented here is of direct experimental relevance for the operational determination of the gravitational field in General Relativity by means of clocks. 

In particular, we extended the results from \cite{Puetzfeld:etal:2018:2} in two ways. First we derived new analytical expressions for the acceleration and angular velocity of the reference frame in terms of measurable frequency ratios of suitable clock configurations. These exact solutions differ from those in \cite{Puetzfeld:etal:2018:2} by a different state of motion of the central reference clock. Additionally, we presented a set of new analytical expressions which allow for a simultaneous determination of the kinematic properties of the underlying reference frame. Furthermore, a new analytical compass solution for all curvature components in Fermi coordinates was obtained. This solution was subsequently classified by the number of actual clock measurements which are required for the determination of each curvature component in the solution. Using this solution, we discussed different experimental strategies to measure particular curvature components. In general the components can either be determined directly/simultaneously -- together with other curvature components from a larger clock configuration -- and/or hierarchically, i.e.\ using the knowledge of previously determined curvature components.

In the second half of our work we illustrated how the statistical determination of some representative curvature components could be carried out. Starting from mock data -- which takes into account possible variability of the measured position and velocity of the clocks, as well as of the corresponding frequency ratios -- we computed the posterior probability distributions of several curvature components by using each of the different approaches (direct/simultaneous, hierarchical). This lead to an estimate of the precision with which each curvature component could be determined in a realistic measurement, and how the resulting probability distribution depends on the various parameters of our model. Some curvature components are better determined by particular clock configurations, depending on the positions/distances, velocities, and the precision of the involved clocks. This behavior was expected, as becomes clear from a comparison to our exact solution, since some of the parameters contribute with different weights to the measured frequency ratio, e.g.\ with factors linear in the velocities, and some with quadratic terms, etc. 

Our results indicate that the strategy of a hierarchical determination of the curvature components leads to an estimation of the curvature of similar precision, using the same data, when compared to the simultaneous approach. Our discussion of the relationship between the different curvature components, and the various alternatives to infer their values from the measurable quantities, is of direct relevance for the future experimental implementation of a clock compass.

It is straightforward to extend our current analysis to include the simultaneous and/or the hierarchical determination of more components of the curvature tensor. Even the full determination of all 20 independent components does not require conceptually different techniques than the ones presented here. By using the model defined by the master Eq.\ \eqref{general_freq_ratio_definition_our}, and suitable position and velocity data of a swarm of clocks as well as their corresponding frequency ratio w.r.t.\ a central clock, all 20 components can be inferred analogously to the examples presented here.

Finally, it should be mentioned that highly accurate clock networks, which are currently in use \cite{Lisdat:etal:2016,BACON:2020} and under construction \cite{Riehle:2017,Bauch:2019}, present an exciting direct application of the framework presented here.

\begin{acknowledgments}
This work was funded by the Deutsche Forschungsgemeinschaft (DFG,  German  Research  Foundation) through  the  grant  PU  461/1-2  –  project  number 369402949 (D.P.), and by the Agencia Nacional de Investigaci\'{o}n y Desarrollo de Chile (ANID, National Agency for Research and Development) /  Scholarship Program / MAGISTER NACIONAL / 2018 - 22182173 (G.N.).
\end{acknowledgments}

\onecolumngrid

\twocolumngrid

\appendix

\section{Notations and conventions}\label{sec_notation}

We follow the notation used in \cite{Puetzfeld:etal:2018:2}, in particular we set $c=1$, raise and lower three dimensional indices of kinematic quantities by means of the Euclidean metric, i.e.\ $\omega^\alpha=\delta^{\alpha\beta}\omega_\beta$, etc. Note however, that for the curvature components we use the convention where the indices are lowered using the full Lorentzian metric, which in turn introduces a different sign in the terms involving $R_{\gamma \alpha \beta \delta}$, c.f.\ Eq.\ \eqref{general_freq_ratio_definition}. The symbols used in this work are summarized in table \ref{tab_symbols}.

\begin{table}
\caption{\label{tab_symbols}Directory of symbols.}
\begin{ruledtabular}
\begin{tabular}{ll}
Symbol & Explanation\\
\hline
$g_{a b}$ & Metric\\
$Y(s)$, $X(\tau)$ & (Reference) world line\\
$\varepsilon_{\alpha\beta\gamma}$  & 3D Levi-Civita symbol\\
$x^{\alpha}$ & Spatial Fermi coordinates \\
$\tau$ & Proper time \\
$\delta_{\alpha\beta}$ & 3D Euclidean metric \\
$R_{abcd}=R_{a b c}{}^eg_{de}$ & Riemann curvature \\
$v^\alpha$, $\omega^\alpha$   & (Linear, angular) velocity\\
$a^\alpha$ & Acceleration\\
$\bar{C}, B_{1,2,3}$ & Auxiliary quantities
\end{tabular}
\end{ruledtabular}
\end{table}

\section{Fully resolved form of the curvature solution}\label{RB}

\begin{align}
  R_{0110} &= - \frac{\pre{B_3}{1,0}}{y^{2}}, \label{R0110B}\\
  R_{0220} &= - \frac{\pre{B_3}{2,0}}{y^{2}}, \label{R0220B}\\
  R_{0330} &= - \frac{\pre{B_3}{3,0}}{y^{2}},  \\
  R_{0120} &= \frac{1}{2 y^{2}} \left(\pre{B_3}{1,0} + \pre{B_3}{2,0} - \pre{B_3}{4,0} \right), \label{R0120B}\\
  R_{0130} &= \frac{1}{2 y^{2}} \left(\pre{B_3}{1,0} + \pre{B_3}{3,0} - \pre{B_3}{6,0} \right), \label{R0130B}\\
  R_{0230} &= \frac{1}{2 y^{2}} \left(\pre{B_3}{2,0} + \pre{B_3}{3,0} - \pre{B_3}{5,0} \right) \label{R0230B}, \\
  R_{1210} &= \frac{3}{8 v_{2 2} y^{2}} \left(\pre{B_3}{1,2} - \pre{B_3}{1,-2}\right), \label{R1210B}\\ 
  R_{1310} &=  \frac{3}{8 v_{33} y^{2}} \left(\pre{B_3}{1,3} - \pre{B_3}{1,-3}\right), \\ 
  R_{2320} &= \frac{3}{8 v_{33} y^{2}} \left(\pre{B_3}{2,3} - \pre{B_3}{2,-3}\right), \\ 
  R_{1212} &= -\frac{3}{2 v^2_{2 2}  y^{2} } \left(2 \pre{B_3}{1,0} - \pre{B_3}{1,2} - \pre{B_3}{1,-2} \right), \label{R1212B}\\ 
  R_{1313} &=  -\frac{3}{2 v^2_{33}  y^{2} } \left(2 \pre{B_3}{1,0} - \pre{B_3}{1,3} - \pre{B_3}{1,-3} \right),  \\ 
  R_{2323} &= -\frac{3}{2 v^2_{33}  y^{2} } \left(2 \pre{B_3}{2,0} - \pre{B_3}{2,3} - \pre{B_3}{2,-3} \right), \\
	R_{1220} &= \frac{3}{8 v_{1 1} v_{22}^2 y^{2}} \left( 2 v_{22}^2 (\pre{B_3}{2,0} - \pre{B_3}{2,1} ) \right. \nonumber \\
	&\left.+ v_{11}^2 ( -2 \pre{B_3}{1,0} + \pre{B_3}{1,2} + \pre{B_3}{1,-2} )   \right),\label{R1220B} \\
  R_{1330} &= \frac{3 }{8 v_{1 1} v_{22}^2 y^{2}} \left( 2 v_{33}^2 (\pre{B_3}{3,0} - \pre{B_3}{3,1} ) \right. \nonumber \\
	&\left.+ v_{11}^2 ( -2 \pre{B_3}{1,0} + \pre{B_3}{1,3} + \pre{B_3}{1,-3} )   \right),\\ 
	R_{2330} &= \frac{3}{8 v_{2 2} v_{33}^2 y^{2}} \left( 2 v_{33}^2 (\pre{B_3}{3,0} - \pre{B_3}{3,1} ) \right. \nonumber \\
	&\left.+ v_{22}^2 ( -2 \pre{B_3}{2,0} + \pre{B_3}{2,3} + \pre{B_3}{2,-3} )   \right), \label{R2330B}
\end{align}

\begin{widetext}
\begin{align}
 R_{1213} &= - \frac{3}{4 v_{2 2}^{2} v_{3 3}^{2} v_{5 2} v_{5 3} y_{0}^{2}} \Big(2 \left(v_{2 2}^{2} v_{3 3}^{2} - v_{2 2}^{2} v_{5 3}^{2} -  v_{3 3}^{2} v_{5 2}^{2} \right) \pre{B_3}{1,0} + \left( v_{2 2} v_{3 3}^{2} v_{5 2} +  v_{3 3}^{2} v_{5 2}^{2} \right)\pre{B_3}{1,2}  \nonumber \\
 &  \quad + \left( v_{2 2}^{2} v_{3 3} v_{5 3} +  v_{2 2}^{2} v_{5 3}^{2}\right) \pre{B_3}{1,3} - 2  v_{2 2}^{2} v_{3 3}^{2} \pre{B_3}{1,5} - \left( v_{2 2} v_{3 3}^{2} v_{5 2}- v_{3 3}^{2} v_{5 2}^{2}\right)\pre{B_3}{1,-2}\nonumber\\
 &\qquad - \left( v_{2 2}^{2} v_{3 3} v_{5 3} - v_{2 2}^{2} v_{5 3}^{2} \right) \pre{B_3}{1,-3}\Big), \label{R1213B} \\ 
 R_{1223} &= \frac{3}{4 v_{1 1} v_{2 2}^{2} v_{3 3}^{2} v_{6 1} v_{6 3} y_{0}^{2}} \Big[2 ( v_{1 1}^{2} v_{3 3}^{2} v_{6 1} - v_{1 1} v_{3 3}^{2} v_{6 1}^{2})\pre{B_3}{1,0} - ( v_{1 1}^{2} v_{3 3}^{2} v_{6 1} - v_{1 1} v_{3 3}^{2} v_{6 1}^{2} ) \pre{B_3}{1,2} \nonumber \\
 & \quad + 2 ( v_{1 1} v_{2 2}^{2} v_{3 3}^{2} - v_{1 1} v_{2 2}^{2} v_{6 3}^{2} -  v_{2 2}^{2} v_{3 3}^{2} v_{6 1} )\pre{B_3}{2,0} + 2 v_{2 2}^{2} v_{3 3}^{2} v_{6 1} \pre{B_3}{2,1} + ( v_{1 1} v_{2 2}^{2} v_{3 3} v_{6 3} +  v_{1 1} v_{2 2}^{2} v_{6 3}^{2} ) \pre{B_3}{2,3} \nonumber \\
 & \quad - 2  v_{1 1} v_{2 2}^{2} v_{3 3}^{2} \pre{B_3}{2,6} - ( v_{1 1}^{2} v_{3 3}^{2} v_{6 1} - v_{1 1} v_{3 3}^{2} v_{6 1}^{2}) \pre{B_3}{1,-2} - v_{1 1} v_{2 2}^{2} v_{3 3} v_{6 3}- v_{1 1} v_{2 2}^{2} v_{6 3}^{2} ) \pre{B_3}{2,-3} \Big], \label{R1223B}\\
 R_{1323} &= - \frac{3}{4 v_{1 1} v_{2 2} v_{3 3}^{2} v_{4 1} v_{4 2} y_{0}^{2}} \Big[2 ( v_{1 1}^{2} v_{2 2} v_{4 1} - v_{1 1} v_{2 2} v_{4 1}^{2})\pre{B_3}{1,0} - ( v_{1 1}^{2} v_{2 2} v_{4 1} - v_{1 1} v_{2 2} v_{4 1}^{2})\pre{B_3}{1,3} \nonumber \\
 & \quad + 2 ( v_{1 1} v_{2 2}^{2} v_{4 2} - v_{1 1} v_{2 2} v_{4 2}^{2} ) \pre{B_3}{2,0} - (v_{1 1} v_{2 2}^{2} v_{4 2}  - v_{1 1} v_{2 2} v_{4 2}^{2} ) \pre{B_3}{2,3} \nonumber \\
 & \quad + 2 ( v_{1 1} v_{2 2} v_{3 3}^{2} -  v_{1 1} v_{3 3}^{2} v_{4 2} - v_{2 2} v_{3 3}^{2} v_{4 1} ) \pre{B_3}{3,0} + 2  v_{2 2} v_{3 3}^{2} v_{4 1} \pre{B_3}{3,1} + 2  v_{1 1} v_{3 3}^{2} v_{4 2} \pre{B_3}{3,2} - 2  v_{1 1} v_{2 2} v_{3 3}^{2} \pre{B_3}{3,4} \nonumber\\ 
 & \quad - ( v_{1 1}^{2} v_{2 2} v_{4 1} - v_{1 1} v_{2 2} v_{4 1}^{2}) \pre{B_3}{1,-3} - ( v_{1 1} v_{2 2}^{2} v_{4 2} - v_{1 1} v_{2 2} v_{4 2}^{2} ) \pre{B_3}{2,-3}\Big], \label{R1323B}\\
  R_{1230} &= - \frac{3}{8 v_{1 1} v_{2 2} v_{3 3} v_{4 1} v_{4 2} y_{0}^{2}} \Big[(2 ( v_{1 1}^{2} v_{2 2} v_{4 1} - v_{1 1} v_{2 2} v_{4 1}^{2} + v_{1 1} v_{2 2} v_{4 1} v_{4 2} ) \pre{B_3}{1,0} \nonumber \\
  & \quad - ( v_{1 1}^{2} v_{2 2} v_{4 1} - v_{1 1} v_{2 2} v_{4 1}^{2} + 2 v_{1 1} v_{2 2} v_{4 1} v_{4 2} ) \pre{B_3}{1,3} + 2 ( v_{1 1} v_{2 2}^{2} v_{4 2} + v_{1 1} v_{2 2} v_{4 1} v_{4 2} -  v_{1 1} v_{2 2} v_{4 2}^{2} ) \pre{B_3}{2,0} \nonumber \\
  &\quad - ( v_{1 1} v_{2 2}^{2} v_{4 2} + 2  v_{1 1} v_{2 2} v_{4 1} v_{4 2} - v_{1 1} v_{2 2} v_{4 2}^{2} ) \pre{B_3}{2,3} + 2 ( v_{1 1} v_{2 2} v_{3 3}^{2} -  v_{1 1} v_{3 3}^{2} v_{4 2} -  v_{2 2} v_{3 3}^{2} v_{4 1} ) \pre{B_3}{3,0} \nonumber \\
  & \quad + 2  v_{2 2} v_{3 3}^{2} v_{4 1} \pre{B_3}{3,1} + 2  v_{1 1} v_{3 3}^{2} v_{4 2} \pre{B_3}{3,2} - 2  v_{1 1} v_{2 2} v_{3 3}^{2} \pre{B_3}{3,4} - 2  v_{1 1} v_{2 2} v_{4 1} v_{4 2} \pre{B_3}{4,0} + 2  v_{1 1} v_{2 2} v_{4 1} v_{4 2} \pre{B_3}{4,3} \nonumber \\
  & \quad -( v_{1 1}^{2} v_{2 2} v_{4 1} - v_{1 1} v_{2 2} v_{4 1}^{2} )  \pre{B_3}{1,-3} - (v_{1 1} v_{2 2}^{2} v_{4 2} -  v_{1 1} v_{2 2} v_{4 2}^{2} ) \pre{B_3}{2,-3} \Big], \\
 R_{2310} &=  \frac{3}{8 v_{1 1} v_{2 2}^{2} v_{3 3}^{2} v_{5 2} v_{5 3} y_{0}^{2}} \Big[2 ( v_{1 1}^{2} v_{2 2}^{2} v_{3 3}^{2} -  v_{1 1}^{2} v_{2 2}^{2} v_{5 3}^{2} - v_{1 1}^{2} v_{3 3}^{2} v_{5 2}^{2} ) \pre{B_3}{1,0} + ( v_{1 1}^{2} v_{2 2} v_{3 3}^{2} v_{5 2} +  v_{1 1}^{2} v_{3 3}^{2} v_{5 2}^{2} ) \pre{B_3}{1,2} \nonumber \\
 & \quad  + ( v_{1 1}^{2} v_{2 2}^{2} v_{3 3} v_{5 3} +  v_{1 1}^{2} v_{2 2}^{2} v_{5 3}^{2} ) \pre{B_3}{1,3} - 2  v_{1 1}^{2} v_{2 2}^{2} v_{3 3}^{2} \pre{B_3}{1,5} + 2  v_{2 2}^{2} v_{3 3}^{2} v_{5 2} v_{5 3} \pre{B_3}{2,0} - 2  v_{2 2}^{2} v_{3 3}^{2} v_{5 2} v_{5 3} \pre{B_3}{2,1} \nonumber \\
 &\quad  + 2  v_{2 2}^{2} v_{3 3}^{2} v_{5 2} v_{5 3} \pre{B_3}{3,0} - 2  v_{2 2}^{2} v_{3 3}^{2} v_{5 2} v_{5 3} \pre{B_3}{3,1} - 2  v_{2 2}^{2} v_{3 3}^{2} v_{5 2} v_{5 3} \pre{B_3}{5,0} + 2  v_{2 2}^{2} v_{3 3}^{2} v_{5 2} v_{5 3} \pre{B_3}{5,1} \nonumber \\
 &\quad - ( v_{1 1}^{2} v_{2 2} v_{3 3}^{2} v_{5 2} - v_{1 1}^{2} v_{3 3}^{2} v_{5 2}^{2}) \pre{B_3}{1,-2} - ( v_{1 1}^{2} v_{2 2}^{2} v_{3 3} v_{5 3} - v_{1 1}^{2} v_{2 2}^{2} v_{5 3}^{2}) \pre{B_3}{1,-3}\Big].  \label{R2310B}
\end{align}
\end{widetext}

\section{Analytical expressions for error propagation}\label{app_RB}

We derive an approximate analytical expression for the error of the curvature. This result is then used to place upper limits, depending on the desired target error for the curvature, on the error of the variables which enter the expression for the curvature. The simple analytical result is useful for the adjustment of parameters in our simulations. 

The general form of a curvature component like $R=R_{0110}$ is of the generic form $R=-(\C+v^2)/y^2$, see \eqref{form-of-B}, therefore, we infer that
\begin{align}
\left( \dfrac{\sigma_{R}}{R} \right)^2 \approx \left( \dfrac{\sigma_{\bar{C}}}{\bar{C} + v^2} \right)^2 + \left( \dfrac{ 2 \sigma_{y}}{y} \right)^2 + \left( \dfrac{ 2 v \sigma_{v}}{\C + v^2} \right)^2 . \label{apendix_C_1}
\end{align}
If we want to have a fractional error lower than a certain value, this expression becomes an inequality,
\begin{equation}
 \left(\frac{\sigma_{\bar{C}}}{\bar{C} + v^2}\right)^2  + \left(\frac{2 \sigma_y }{y}\right)^2 + \left(\frac{2 v\sigma_v }{\bar{C} + v^2}\right)^2 < \left( \dfrac{\sigma_R}{R} \right)^2, 
\end{equation}
which requires that
\begin{align}
 \left(\frac{\sigma_{\bar{C}}}{\bar{C} + v^2}\right)^2 &< \left( \dfrac{\sigma_R}{R} \right)^2, \\
 \left(\frac{2 \sigma_y }{y}\right)^2 &< \left( \dfrac{\sigma_R}{R} \right)^2,
\intertext{and}
\left(\frac{2 v\sigma_v }{\bar{C} + v^2}\right)^2 &< \left( \dfrac{\sigma_R}{R} \right)^2.
\end{align}
From this, we derive the necessary conditions
\begin{align}
\sigma_{\bar{C}} &< \left| \frac{\sigma_R}{R} \right| |\bar{C} + v^2|, \label{pred_C} \\
\sigma_y  &< \left| \frac{\sigma_R}{R} \right| \dfrac{y}{2}, \label{pred_y} 
\intertext{and}
\sigma_v  &< \left| \frac{\sigma_R}{R} \dfrac{\bar{C} + v^2}{2v} \right|. \label{pred_v} 
\end{align}

\bibliographystyle{unsrtnat}
\bibliography{extcomp_bibliography}

\begin{thebibliography}{19}
\providecommand{\natexlab}[1]{#1}
\providecommand{\url}[1]{\texttt{#1}}
\expandafter\ifx\csname urlstyle\endcsname\relax
  \providecommand{\doi}[1]{doi: #1}\else
  \providecommand{\doi}{doi: \begingroup \urlstyle{rm}\Url}\fi

\bibitem[{Puetzfeld} et~al.(2018){Puetzfeld}, {Obukhov}, and
  {L\"ammerzahl}]{Puetzfeld:etal:2018:2}
D.~{Puetzfeld}, Y.~N. {Obukhov}, and C.~{L\"ammerzahl}.
\newblock {Gravitational clock compass in General Relativity}.
\newblock \emph{Phys. Rev. D}, 98:\penalty0 024032, 2018.
\newblock \doi{10.1103/PhysRevD.98.024032}.

\bibitem[{Szekeres}(1965)]{Szekeres:1965}
P.~{Szekeres}.
\newblock {The gravitational compass}.
\newblock \emph{J. Math. Phys.}, 6:\penalty0 1387, 1965.
\newblock \doi{10.1063/1.1704788}.

\bibitem[{Pirani}(1956)]{Pirani:1956}
F.~A.~E. {Pirani}.
\newblock {On the physical significance of the Riemann tensor}.
\newblock \emph{Acta Phys. Pol.}, 15:\penalty0 389, 1956.
\newblock \doi{10.1007/s10714-009-0787-9}.

\bibitem[{Synge}(1960)]{Synge:1960}
J.~L. {Synge}.
\newblock \emph{{Relativity: The general theory}}.
\newblock North-Holland, Amsterdam, 1960.

\bibitem[{Obukhov} and {Puetzfeld}(2019)]{Obukhov:Puetzfeld:2019}
Y.~N. {Obukhov} and D.~{Puetzfeld}.
\newblock {Measuring the gravitational field in General Relativity: From
  deviation equations and the gravitational compass to relativistic clock
  gradiometry}.
\newblock \emph{``Relativistic Geodesy: Foundations and Applications'', D.\
  Puetzfeld et. al. (eds.), Fundamental Theories of Physics, Springer (Cham)},
  196:\penalty0 87, 2019.
\newblock \doi{10.1007/978-3-030-11500-5_3}.

\bibitem[{Puetzfeld} and {Obukhov}(2016)]{Puetzfeld:Obukhov:2016:1}
D.~{Puetzfeld} and Y.~N. {Obukhov}.
\newblock {Generalized deviation equation and determination of the curvature in
  General Relativity}.
\newblock \emph{Phys. Rev. D}, 93:\penalty0 044073, 2016.
\newblock \doi{10.1103/PhysRevD.93.044073}.

\bibitem[{Hogan} and {Puetzfeld}(2020)]{Hogan:Puetzfeld:2020:1}
P.~A. {Hogan} and D.~{Puetzfeld}.
\newblock {Gravitational clock compass and the detection of gravitational
  waves}.
\newblock \emph{Phys. Rev. D}, 101:\penalty0 044012, 2020.
\newblock \doi{10.1103/PhysRevD.101.044012}.

\bibitem[{Ni} and {Zimmermann}(1978)]{Ni:Zimmermann:1978}
W.~T. {Ni} and M.~{Zimmermann}.
\newblock {Inertial and gravitational effects in the proper reference frame of
  an accelerated, rotating observer}.
\newblock \emph{Phys. Rev. D}, 17:\penalty0 1473, 1978.
\newblock \doi{10.1103/PhysRevD.17.1473}.

\bibitem[{Le Poncin-Lafitte} et~al.(2004){Le Poncin-Lafitte}, {Linet}, and
  {Teyssandier}]{Poncin-Lafitte:etal:2004}
C.~{Le Poncin-Lafitte}, B.~{Linet}, and P.~{Teyssandier}.
\newblock {World function and time transfer: general post-Minkowskian
  expansions}.
\newblock \emph{Class. Quantum Grav.}, 21:\penalty0 4463, 2004.
\newblock \doi{10.1088/0264-9381/21/18/012}.

\bibitem[{Teyssandier} et~al.(2008){Teyssandier}, {Le Poncin-Lafitte}, and
  {Linet}]{Teyssandier:etal:2008:1}
P.~{Teyssandier}, C.~{Le Poncin-Lafitte}, and B.~{Linet}.
\newblock {A universal tool for determining the time delay and the frequency
  shift of light: Synge’s World function}.
\newblock \emph{In ''Lasers, Clocks and Drag-Free Control: Exploration of
  Relativistic Gravity in Space'', H. Dittus, C. L\"{a}mmerzahl, and S. G.
  Turyshev (eds), Springer (Berlin), Astrophysics and Space Science Library},
  349:\penalty0 153, 2008.
\newblock \doi{10.1007/978-3-540-34377-6_6}.

\bibitem[{Teyssandier} and {Le Poncin-Lafitte}(2008)]{Teyssandier:etal:2008:2}
P.~{Teyssandier} and C.~{Le Poncin-Lafitte}.
\newblock {General post-Minkowskian expansion of time transfer functions}.
\newblock \emph{Class. Quantum Grav.}, 25:\penalty0 145020, 2008.
\newblock \doi{10.1088/0264-9381/25/14/145020}.

\bibitem[{Chou} et~al.(2010{\natexlab{a}}){Chou}, {Hume}, {Rosenband}, and
  {Wineland}]{Chou:etal:2010:1}
C.~W. {Chou}, D.~B. {Hume}, T.~{Rosenband}, and D.~J. {Wineland}.
\newblock {Optical clocks and relativity}.
\newblock \emph{Science}, 329:\penalty0 1630, 2010{\natexlab{a}}.
\newblock \doi{10.1126/science.1192720}.

\bibitem[{Chou} et~al.(2010{\natexlab{b}}){Chou}, {Hume}, {Koelemeij},
  {Wineland}, and {Rosenband}]{Chou:etal:2010:2}
C.~W. {Chou}, D.~B. {Hume}, J.~C.~J. {Koelemeij}, D.~J. {Wineland}, and
  T.~{Rosenband}.
\newblock {Frequency comparison of two high-accuracy Al$^+$ optical clocks}.
\newblock \emph{Phys. Rev. Lett.}, 104:\penalty0 070802, 2010{\natexlab{b}}.
\newblock \doi{10.1103/PhysRevLett.104.070802}.

\bibitem[{Foreman-Mackey} et~al.(2013){Foreman-Mackey}, {Hogg}, {Lang}, and
  {Goodman}]{emcee:2013}
D.~{Foreman-Mackey}, D.~W. {Hogg}, D.~{Lang}, and J.~{Goodman}.
\newblock {emcee: The MCMC Hammer}.
\newblock \emph{PASP}, 125:\penalty0 306, 2013.
\newblock \doi{10.1086/670067}.

\bibitem[{Bini} et~al.(2005){Bini}, {Geralico}, and {Jantzen}]{Bini:2005}
D.~{Bini}, A.~{Geralico}, and R.~T. {Jantzen}.
\newblock {Kerr metric, static observers and Fermi coordinates}.
\newblock \emph{Class. Quant. Grav.}, 22:\penalty0 4729, 2005.
\newblock \doi{10.1088/0264-9381/22/22/006}.

\bibitem[{Lisdat} and et~al.(2016)]{Lisdat:etal:2016}
C.~{Lisdat} and et~al.
\newblock {A clock network for geodesy and fundamental science}.
\newblock \emph{Nature Communications}, 7:\penalty0 12443, 2016.
\newblock \doi{10.1038/ncomms12443}.

\bibitem[Collaboration(2020)]{BACON:2020}
Boulder Atomic Clock Optical Network~(BACON) Collaboration.
\newblock {Frequency ratio measurements with 18-digit accuracy using a network
  of optical clocks}.
\newblock \emph{arXiv:2005.14694 [physics.atom-ph]}, 2020.
\newblock URL \url{https://arxiv.org/abs/2005.14694}.

\bibitem[{Riehle}(2017)]{Riehle:2017}
F.~{Riehle}.
\newblock {Optical clock networks}.
\newblock \emph{Nature Photonics}, 11:\penalty0 25, 2017.
\newblock \doi{10.1038/nphoton.2016.235}.

\bibitem[{Bauch}(2019)]{Bauch:2019}
A.~{Bauch}.
\newblock {Time and frequency metrology in the context of relativistic
  geodesy}.
\newblock \emph{``Relativistic Geodesy: Foundations and Applications'', D.\
  Puetzfeld et. al. (eds.), Fundamental Theories of Physics, Springer (Cham)},
  196:\penalty0 1, 2019.
\newblock \doi{10.1007/978-3-030-11500-5_1}.

\end{thebibliography}
\end{document}